\documentclass[lettersize,journal]{IEEEtran}
\usepackage{array}
\usepackage[caption=false,font=normalsize,labelfont=sf,textfont=sf]{subfig}
\usepackage{textcomp}
\usepackage{stfloats}
\usepackage{xcolor}
\usepackage{url}
\usepackage{verbatim}
\usepackage{graphicx}

\usepackage{amsmath,amssymb,amsfonts}
\usepackage{algorithm}
\usepackage{amsmath}
\usepackage{algorithmicx}
\usepackage{algpseudocode}
\usepackage{textcomp}
\usepackage{xcolor}
\usepackage{multirow}
\usepackage{bm}
\usepackage{makecell}
\usepackage{booktabs}
\usepackage{ntheorem}
\usepackage{tabularx}
\usepackage{booktabs}

\def\BibTeX{{\rm B\kern-.05em{\sc i\kern-.025em b}\kern-.08em
 T\kern-.1667em\lower.7ex\hbox{E}\kern-.125emX}}
\usepackage{balance}
\begin{document}
	
\title{Multi-objective Optimization for Data Collection in UAV-assisted Agricultural IoT}

\author{Lingling Liu,
	Aimin Wang, Geng~Sun,~\IEEEmembership{Member,~IEEE},
	Jiahui~Li,~\IEEEmembership{Student Member,~IEEE}, \\
	Hongyang Pan, and Tony Q. S. Quek,~\IEEEmembership{Fellow,~IEEE}
\thanks{This research is supported in part by the National Natural Science Foundation of China (62172186, 62272194, 61872158, 62002133); in part by the Science and Technology Development Plan Project of Jilin Province (20230201087GX); in part by the Young Science and Technology Talent Lift Project of Jilin Province (QT202013); in part by the Graduate Innovation Fund of Jilin University (2022110); and in part by China Scholarship Council (\textit{Corresponding author: Geng Sun}).\protect}
\thanks{Lingling Liu is with the College of Computer Science and Technology, Jilin University, Changchun 130012, China, and also with the Information Systems Technology and Design (ISTD) Pillar, Singapore University of Technology and Design, Singapore 487372 (e-mail: linglingliu2020@foxmail.com).} 

\thanks{Aimin Wang and Geng Sun are with the College of Computer Science and Technology, Jilin University, Changchun 130012, China (e-mail: wangam@jlu.edu.cn; sungeng@jlu.edu.cn).}

\thanks{Jiahui Li is with the College of Computer Science and Technology, Jilin University, Changchun 130012, China, and also with Pillar of Engineering Systems and Design (ESD), Singapore University of Technology and Design, Singapore 487372 (e-mail: lijiahui0803@foxmail.com).}

\thanks{Hongyang Pan is with the College of Computer Science and Technology, Jilin University, Changchun 130012, China, and also with the Engineering Product Development (EPD) Pillar, Singapore University of Technology and Design, Singapore 487372 (e-mail: panhongyang18@foxmail.com}

\thanks{Tony Q. S. Quek is with the Singapore University of Technology and Design, Singapore 487372 (e-mail: tonyquek@sutd.edu.sg).}}

\markboth{Journal of \LaTeX\ Class Files,~Vol.~18, No.~9, September~2022}%
{How to Use the IEEEtran \LaTeX \ Templates}

\maketitle

\begin{abstract} 
The ground fixed base stations (BSs) are often deployed inflexibly, and have high overheads, as well as are susceptible to the damage from natural disasters, making it impractical for them to continuously collect data from sensor devices. To improve the network coverage and performance of wireless communication, unmanned aerial vehicles (UAVs) have been introduced in diverse wireless networks, therefore in this work we consider employing a UAV as an aerial BS to acquire data of agricultural Internet of Things (IoT) devices. To this end, we first formulate a UAV-assisted data collection multi-objective optimization problem (UDCMOP) to efficiently collect the data from agricultural sensing devices. Specifically, we aim to collaboratively optimize the hovering positions of UAV, visit sequence of UAV, speed of UAV, in addition to the transmit power of devices, to simultaneously achieve the maximization of minimum transmit rate of devices, the minimization of total energy consumption of devices, and the minimization of total energy consumption of UAV. Second, the proposed UDCMOP is a non-convex mixed integer nonlinear optimization problem, which indicates that it includes continuous and discrete solutions, making it intractable to be solved. Therefore, we solve it by proposing an improved multi-objective artificial hummingbird algorithm (IMOAHA) with several specific improvement factors, that are the hybrid initialization operator, Cauchy mutation foraging operator, in addition to the discrete mutation operator. Finally, simulations are carried out to testify that the proposed IMOAHA can effectively improve the system performance comparing to other benchmarks.
\end{abstract}


\section{Introduction}
\IEEEPARstart{T}{he} Internet of Things (IoT) is a vital technology of contemporary wireless communication systems~\cite{park2021low}, and has been widely applied in diverse fields, such as intelligent cities~\cite{sanchez2014smartsantander}, smart homes~\cite{park2017comprehensive}, and smart agriculture~\cite{zhou2021UAV}. Among them, the IoT in agriculture has developed rapidly and plays a key role in modern agricultural production. Furthermore, the rapid implement of IoT tightly depends on the development of wireless sensors networks (WSNs)~\cite{park2021low}, which are made up of massive sensors with the abilities of sensing, processing and communication. However, due to the constraints of technology and finance, sensors are battery-powered devices, and their limited lifetime has a great effect on the performance improvements and perception ability in WSNs, which results in a lot of researches devoted to increasing the lifetime of sensors~\cite{tan2019adaptive},~\cite{mostafaei2018energy}. Nevertheless, the storage capacity of sensor is usually limited, and most of them need to upload data to base stations (BSs) or sink nodes through multiple hops, which may lead to lower transmit rates and even data loss.

\par Due to their constantly dropping costs, growing functionality, and the ability to introduce appealing applications, unmanned aerial vehicles (UAVs) are popularly used in numerous significant wireless communication applications~\cite{zeng2016wireless}. Furthermore, they have the capacities of low-cost detection and data analysis, which enables them to be used as aerial base stations (BSs) for collecting data~\cite{duan2019resource} and relaying data~\cite{ji2019performance}. Moreover, the application of UAVs in agriculture can greatly reduce human resources, and the data collected by UAVs can be timely evaluated and utilized, which is also conducive to making accurate judgments and improving crop production~\cite{aslan2022comprehensive}, efficiently manage crops~\cite{lopez2021framework}. Inspired by the abovementioned descriptions, applying UAVs into agriculture to observe crop growth status has attracted great research interest in both academic and industrial fields. 

\par Nevertheless, in existing UAV-assisted agricultural IoT data collection scenarios, UAV can obtain data from multiple sensor devices at the same time. On the one hand, this will allow devices closer to UAV to transmit data at an optimal transmit rate, while the transmit rate is relatively lower for sensor devices that are far away. On the other hand, these sensor devices are usually not rechargeable or charged by a power resource, which means that the devices cannot work continuously. Therefore, it is crucial to upload almost all data to UAV before the energy is exhausted. For the purpose of solving the abovementioned challenges, UAV can hover at different optimal positions to collect data from the covered sensor devices by optimizing their movement paths and hovering locations. In addition, UAVs will consume nonnegligible energy in the process of changing positions and collecting data.

\par In contrast to earlier works about agricultural IoT, in which the energy consumed by sensor devices, the transmit rates of devices, or the energy consumption of UAVs are all taken into account unilaterally, we comprehensively take these three indexes as the optimization objectives. Moreover, the optimization variables of our proposed problem include the locations of UAV, speed of UAV, transmit power of sensor devices, and the visit sequence of UAV, which are respectively continuous and discrete variables, making it intractable to be tackled. The multi-objective evolutionary algorithms (MOEAs) have some advantages in solving multi-objective optimization problems (MOPs). For example, it does not need to divide a MOP into multiple single objective subproblems, or convert it into a single objective problem through multiple conversions. Moreover, it does not require complex and sufficient computational theoretical knowledge, therefore we adopt the MOEAs to solve the abovementioned MOPs. Following is a list of the main contributions of this work:

\begin{itemize}

	\item We consider such a scenario that a UAV starts from a given initial location, moves through each agricultural IoT subarea, then collects data from sensor devices in the area, and finally uploads data to BS at a given end location. Moreover, a UAV-assisted data collection multi-objective optimization problem (UDCMOP) in agricultural IoT is formulated to simultaneously maximize the minimum transmit rate of sensor devices, minimize the total energy consumption of devices, and minimize the total energy consumption of UAV by cooperatively optimizing the positions of UAV, the transmit power of sensor devices, speed of UAV, in addition to the visit sequences of UAV.
	
	\item In the solutions of the formulated UDCMOP, the positions and speed of UAV, and transmit power of sensor devices are all continuous solution components, while the visit sequence of UAV is discrete solution component, causing it intractable to be solved. Thereupon, an improved multi-objective artificial hummingbird algorithm (IMOAHA) is proposed to deal with the problem. Specifically, IMOAHA adopts a hybrid initialization operator to improve the diversity and quality of the solutions. Then, a Cauchy mutation foraging operator is used to enhance the global search capability of IMOAHA. Moreover, a discrete solution update operator is utilized to enhance the searching efficiency of discrete solutions, so that balancing the exploration and exploitation abilities of the proposed IMOAHA. The abovementioned designs make IMOAHA appropriate for dealing with the formulated UDCMOP.
	
	\item Comprehensive simulations are executed to assess the performance and effectiveness of the suggested IMOAHA, and numerical outcomes show that the suggested scheme simultaneously increase transmit rate, and decrease energy consumption of sensor devices and UAV comparing to other methods. Furthermore, a series of more intuitive outcomes demonstrate that the proposed IMOAHA is closer to the Pareto Front (PF) than other comparative approaches.

\end{itemize}

\par The rest of this work is organized as follows. Section \ref{Literature review} provides some recent related research progresses. The models and preliminaries are introduced in Section \ref{System Models}. Section \ref{Problem formulation} formulates and analyses UDCMOP in agricultural IoT. The proposed IMOAHA is presented in Section \ref{Methodology}. Section \ref{Simulation results and analysis} shows the results of simulations. Section \ref{Conclusion} is devoted to the conclusions.

\section{Literature review}
\label{Literature review}

\par In this section, the related works are reviewed, including the applications of a UAV in agriculture, UAV-assistant fresh data collection, UAV-assistant energy-efficient data collection, UAV-assistant data collection of rechargeable WSN, and resources allocation of UAV-assisted WSN.

\subsection{Application of UAV in agriculture}

\par Numerous studies and applications about UAV have been widely conducted. In~\cite{radoglou2020compilation}, the authors investigate the application of 20 UAVs in agriculture, including aerial crop detection and spraying. In order to monitor crops of precision agriculture, the authors in~\cite{popescu2020advanced} propose a hierarchical structure based on the cooperation between UAVs and federated WSNs. Liu et al.~\cite{liu2023agricultural} propose a multi-mechanism collaborative improved gray wolf optimization algorithm to solve the issue of agriculture UAV trajectory planning. In~\cite{si2022target}, internet of UAVs is applied to extend the application of insecticidal lamps IoTs (ILs-IoTs), so that the migratory agricultural pests can be quickly killed. For the purpose of addressing the constrained battery capacity of sensor nodes in precision agriculture, a framework for charging sensor nodes and collecting data of sensor nodes by using UAVs is proposed in~\cite{chien2021UAV}. In response to the issue of a large amount of data generated by agricultural production, disaster monitoring, and environmental protection, the authors in~\cite{zhang2021deploying} study the ``air-to-ground" intelligent softwarized collection system to complete the data collection process of IoT nodes. Chen et al.~\cite{chen2021identification} study the application of agricultural UAVs in orchards, in which UAVs only spray pesticides where needed, achieving an increase in crop yield while reducing the use of pesticides.

\subsection{UAV-assisted fresh data collection}

\par The timing of data gathering in current UAV-assisted data collecting methods is essential to maintain the regular operation of system. Therefore, the age of information (AoI) has been used to measure the temporal correlation of the data packets that have been gathered by devices~\cite{sun2021aoi}, in which the propulsion energy consumption of a UAV, average AoI, and the transmission energy consumption of IoT devices are reduced. In addition, in order to look for the ideal UAV trajectory and sensor transmission schedule, Yi et al.~\cite{yi2020deep} investigate the problem of fresh data collection in UAV-assisted IoT networks to minimize the weighted sum of AoI. In~\cite{feng2023age}, the peak age of information (PAoI) of UAV-aided WSNs is reduced by reducing the block size and the queue length. The authors in~\cite{lu2021covertness} study how to maximize the timeliness of data collection under covertness constraints, where UAVs regularly transmit wireless power to charge IoT devices with limited energy, which then send the collected data to the UAVs. 

\subsection{UAV-assisted energy-efficient data collection}

\par In traditional networks, only the ground BSs are utilized to realize the functions of data collection and 3D positioning. On the one hand, the energy consumed by the cell edge devices is higher, and the height of ground BSs is relatively similar, which leads to poor positioning performance of the devices in elevation. To this end, the authors in~\cite{wang2019energy} suggest a new UAV-assisted IoT network that can minimize the maximum energy consumption of all devices. In terms of UAV-assisted WSNs in mountainous locations, a rapid and energy-efficient data gathering system is created with the aid of a UAV acting as a data mule~\cite{nazib2021energy}. Hao et al.~\cite{hao2022joint} study the data acquisition issue in UAV-enabled IoT to minimize the task completion time of UAV while further reducing the energy consumption of UAVs. On the other hand, the task completion time is closely related with the data volume. Zhan et al.~\cite{zhan2019completion} investigate the problem of data collection of a group of sensor nodes in a WSN assisted by multiple UAVs to minimize the maximum task completion time of all UAVs, while making each sensor node successfully upload target data under a given energy budget. 

\subsection{UAV-assisted data collection of rechargeable WSN}

\par Wireless power transmission (WPT) is a promising technology for supplementing sensor devices with limited battery capacity, and the combination of UAVs and WPT technology has been studied in~\cite{yan2020uav}, in which the total energy received by all sensors is maximized to determine the best method for UAV deployment. So as to maximize the remaining energy of UAV in solar powered UAV system while the requirements of IoTDs are satisfied, the scheduling of IoTDs and UAV three-dimensional (3D) trajectory are jointly optimized in~\cite{fu2021joint}. Since UAVs have a finite amount of energy, in order to offer continuous coverage and long-term information services for IoT nodes, the authors in~\cite{li2019rechargeable} investigate energy-efficient cooperative solutions for rechargeable UAVs to maximizes the energy efficiency of the system. In~\cite{luo2020joint}, multiple UAVs have been used in IoT networks to study the data collection issue, in which multiple IoT devices are initially powered by UAVs using WPT, and after that, the powered IoT devices use the energy to upload data to UAVs.

\subsection{Resources allocation of UAV-assisted WSN}

\par The authors in~\cite{fan2023channel} study resource allocation in UAV networks using non-orthogonal multiple access (NOMA) to alleviate the spectrum underutilization problem of single-channel orthogonal multiple access. In~\cite{wang2023secrecy}, the authors use UAVs as relays to maximize secrecy-energy-efficiency. In wireless systems where UAVs serve as aerial BSs to provide downlink data services to ground users, in order to maximize system throughput, wang et al.~\cite{wang2023joint} jointly consider the subchannel assignment, UAV trajectory, and power allocation at per time slot. To maximize the throughput of secondary user (SU), Yu et al.~\cite{yu2023joint} jointly optimize the power allocation of UAV, trajectory of UAV, in addition to the passive beamforming of RIS. In~\cite{wu2023uav}, the impact of UAV jitter on the security performance has been studied by taking into account the secure uplink transmission of analog cooperative beamforming from ground nodes to a UAV receiver. The authors in~\cite{nguyen2021uav} maximize the average achievable secrecy rate of the secondary system by jointly optimize the transmit power and 3D trajectory of UAV. Liu et al.~\cite{liu2022energy} study the dual-UAV-assisted IoT with NOMA, which ensures a certain throughput while reducing UAV energy consumption.

\par Although the energy consumed by the devices, the end-to-end data transmission rate, or the energy consumption of the UAVs is taken into account in the works abovementioned, and did not simultaneously consider these three objectives. Especially in the application of UAVs in agriculture, there is almost no research on trajectory of UAVs for the purposes of data collection. Therefore, our work focuses on this gap, and simultaneously address these three objectives to improve the performance of the system.

\section{System Models and Preliminaries}
\label{System Models}

\par This section will briefly introduce the UAV-assisted system model of agricultural IoT, and the energy consumption models of IoT devices and UAV.
 
\subsection{Scenarios and communication model}

\par As shown in Fig. \ref{fig: Network model}, $K$ IoT devices are placed on the farmland to observe crop growth, soil temperature, and soil moisture, and the farmland is divided into several subarea without overlapping. A UAV is deployed over the monitor area to gather the sensing data from devices, and it is assumed that all sensor nodes are static in our work.

\par Without loss of generality, the 3D coordinate of UAV hovering at position $u$ can be expressed as $\mathcal{U}_u=[x_{u}, y_{u}, z] \in \mathcal{R}^{1 \times 3}$, in which the altitude $z$ of UAV is a fixed constant. Let $\mathcal{S} = \{s_k,1 \leq k \leq K \}$ represent the coordinate set of the IoT devices on the farmland, with $s_k = [x^k, y^k] \in \mathbb{R}^{1 \times 2}$ denoting the horizontal coordinate of device $s_k \in \mathcal{S}$. We divide $K$ devices into $U$ clusters, and the devices in the same cluster upload data to UAV by employing time division multiple access (TDMA) technology to reduce the intra-cluster interference. The cluster can be given by $C_u$, and the set of clusters can be expressed as $C = \{C_1, C_2, ..., C_U\}$. Moreover, it is supposed that the specific locations of these devices are known to UAV in advance, such that it can perform the trajectory planning. Moreover, the distance $d(s_k, u)$ between device $s_k$ and UAV hovering at position $u$ is stated as follows:

\begin{figure}[htbp]
	\centering
	\includegraphics[width=3.4in]{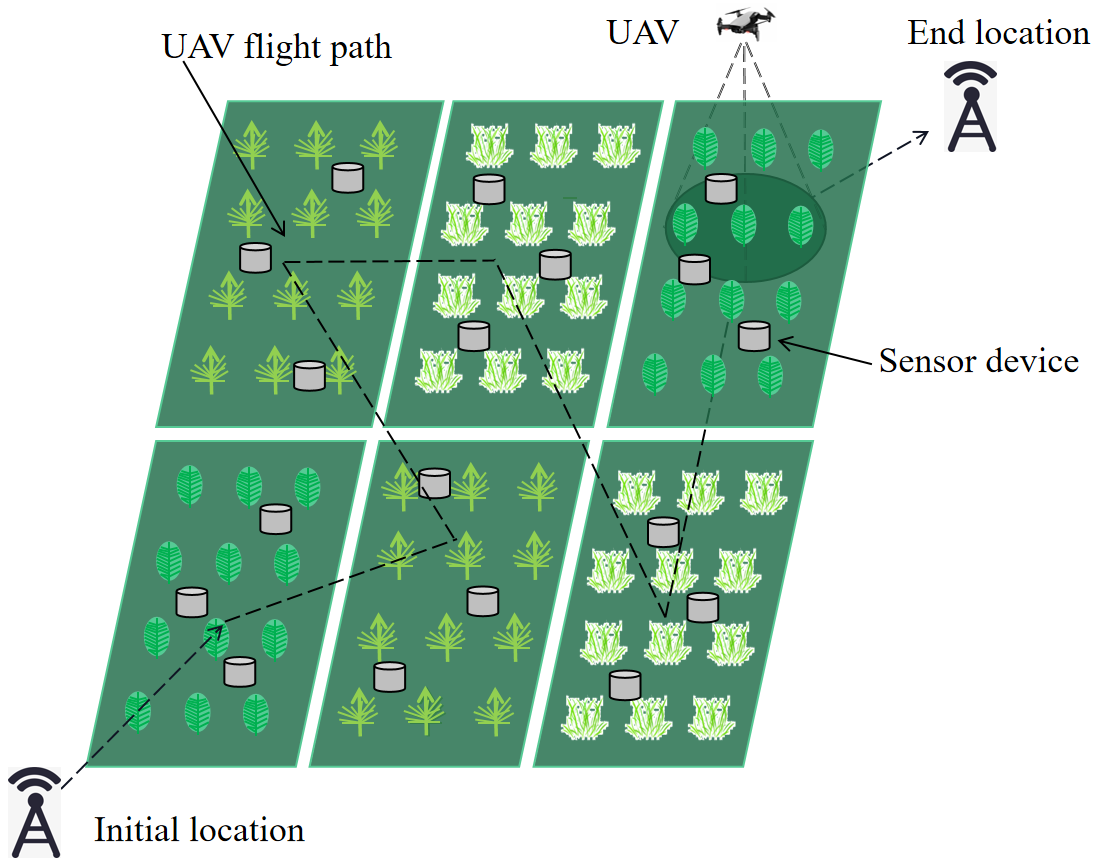}
	\caption{UAV-assisted agricultural IoT model.}
	\label{fig: Network model}
\end{figure}
\begin{equation}
	d{(s_k, u)} = \sqrt{(x_{u} - x^k)^2 + (y_{u} - y^k)^2 + z^2}.
\end{equation} 

\par The data volume of device $s_k$ to be transmitted to the UAV is randomly distributed within the range of ($Q_{\min}, Q_{\max}$), in which $Q_{\min}$ and $Q_{\max}$ are the minimum and maximum data volume accumulated by devices, respectively. In the system under consideration, there may be both line-of-sight (LoS) and non-line-of-sight (NLoS) links between the devices and UAV. With the changing of UAV position, the probability of LoS link may vary. In addition, due to the uncertain obstacles such as taller species, there may still be NLoS link between terrestrial devices and UAV. Correspondingly, the LoS probability from device $s_k$ to UAV hovering at position $u$ is designed as \cite{al2014optimal}.

\begin{equation}
	P_{k, u}^{LoS, \theta} = \frac{1}{1 + C \exp[-B_c(\theta_{k, u}) - C]},
\end{equation}

\noindent where $C$ and $B_c$ denote two constants, which are determined by the environment (such as rural, urban and dense urban) and carrier frequency, respectively. $\theta_{k, u}$ represents the elevation angle from device $s_k$ to UAV at the location $u$, which is computed by $\theta_{k, u} = (180/ \pi ) \arcsin(H/d_{k, u})$. The NLoS probability is denoted by $P_{k, u}^{NLoS, \theta} = 1- P_{k, u}^{LoS, \theta}$. Moreover, the pathloss of LoS and NLoS links can be calculated as \cite{you2020hybrid}:

\begin{equation}
	\label{pathloss}
	h_{k,u}^L = \beta_0 d_{k,u} ^{-\alpha_L},
	h_{k,u}^N = \mu_0 \beta_0 d_{k,u} ^{-\alpha_N},
\end{equation}

\noindent where the additional signal attenuation factor caused by NLoS propagation is represented by $\mu_0$ \cite{you2020hybrid}. $\beta_0$ denotes the average reference channel power gain in LoS link. Therefore, the average pathloss from the device $s_k$ to UAV at position $u$ is given by:

\begin{equation}
	g_{k, u} = P_{k, u}^{LoS, \theta} h_{k,u}^L + \left(1- P_{k, u}^{LoS, \theta}\right)h_{k,u}^N.
\end{equation}

\par Let $p_k$ denote the transmit power of device $s_k$ when it offloads data to UAV. The binary variable $c_{k,u}$ is used to indicate whether the device $s_k$ uploads data to UAV. Specifically, if the UAV hovers over the area $u$ ($u \in C$) where the device $s_k$ belongs, then $c_{k,u} = 1$, otherwise $c_{k,u} = 0$. Therefore, when the device $s_k$ is used to upload data, the transmit rate from device $s_k$ to the UAV deployed at position $u$ can be calculated as \cite{zhan2020completion}:

\begin{equation}
	R_{k, u} =c_{k,u} B \log_{2} \left( 1 + \frac{p_k g_{k, u}}{\sigma^2} \right),
\end{equation}

\noindent where the channel bandwidth is represented as $B$ in Hz, and $\sigma^2$ denotes the additive white Gaussian noise (AWGN) at the $k$th terrestrial IoT device.

\subsection{Energy consumption model of device}

\par We assume that the data sensed by devices will be immediately transmitted to the communicated UAV, and the computing task will not be performed locally on the devices. Therefore, the energy consumption generated by devices computing will not be considered in this work. In addition, the communication energy consumption $E_{k,u}$ of device $k$ is mainly used to upload data to UAV that hovers at position $u$, and the corresponding expression can be designed as:

\begin{equation}
	E_{k, u} = p_kt_{k, u} = p_k \frac{Q_{k, u}}{R_{k, u}} = \mathbb{C} \frac{p_k Q_{k, u}}{\log_{2} \left( 1 + \frac{p_k g_{k,u}}{\sigma^2} \right)},
\end{equation}

\noindent where $Q_{k, u}$ and $t_{k, u}$ represent the data volume to be uploaded by device $s_k$ to UAV hovering at position $u$, and the required time, respectively, in addition to $\mathbb{C}=1/{(c_{k,u} B)}$.

\subsection{Energy consumption model of UAV}
\par The propulsion energy consumption and communication energy consumption make up the majority of energy consumed by UAVs \cite{zeng2019energy}. Specifically, in our considered scenario, the former is used to keep the UAV moving up or forward, while the latter mostly includes communication circuitry, signal processing and signal reception, etc. The energy consumptions during UAV hovering (i.e., data uploading) and UAV movement are designed as follows.

\emph{1) UAV movement phase:} Similar to \cite{zeng2019energy}, we adopt the fly-hover-communication protocol, which indicates that the UAV will not communicate with any device during movement. When UAV moves between multiple hovering points with a fixed speed $V$, the consumed power can be modeled as \cite{zeng2019energy}:

\begin{equation}
	\begin{aligned}
		\label{power consumption}
		\\P\left ( V\right )=&\underset{\textbf{blade\ profile}}{\underbrace{P_{0}\left(1+\frac{3V^{2}}{U_{tip}^{2}}\right)}}+
		\\&\underset{\textbf{induced}}{\underbrace{P_{i}\left ( \sqrt{1+\frac{V^{4}}{4v_{0}^{4}}}-\frac{V^{2}}{2v_{0}^{2}}\right )^{1/2}}}+\underset{\textbf{parasite}}{\underbrace{\frac{1}{2}d_{0}\rho sA_aV^{3}}},
	\end{aligned}
\end{equation}

\noindent where $U_{tip}$ and $v_0$ represent the tip speed of the rotor blade and the mean rotor induced velocity, respectively. $s$ and $d_0$ indicate the rotor solidity and fuselage drag ratio, respectively. Moreover, $A_a$ and $\rho$ respectively denote the rotor disc area and air density. The energy consumed by UAV during its movement can be calculated as follows:

\begin{equation}
	E^m = P(V) \frac{D}{V},
\end{equation}

\noindent where $D$ represents the total path length of UAV movement. It is noted that the speed of a UAV flying from one hovering point to another is fixed, while the speed may vary between two different flight segments. Note that in this article, we ignore the acceleration since it only accounts for a small part of UAV braking process.

\emph{2) UAV hovering phase:} To gather data from the devices, the UAV hovers at a predetermined point for a while, and the consumed energy is calculated by:

\begin{equation}
	E^u = (P_0 + P_i)t_{k, u},
\end{equation}

\noindent where $P_0$ and $P_i$ are obtained when the speed of UAV is set to 0, which are determined by the weight of a UAV, air density and rotor disk area of a UAV~\cite{zeng2019energy}. Moreover, $t_{k, u}$ is the time spent by the UAV in the $u$th hovering position and communicates with the device $s_k$.

\section{Problem formulation and analysis}
\label{Problem formulation and analysis}
\subsection{Problem formulation}
\par Considering a scenario of agricultural IoT shown in Fig. \ref{fig: Network model}, a UAV is dispatched to collect data from $K$ IoT devices of farmland, that are divided into several separate subareas. Since there are diverse channels between the UAV and different devices, the data rate of these devices will be different, resulting in different energy consumption of devices. Moreover, once these devices are installed, they cannot be moved, which means that it is often unrealistic to deploy a UAV only in a fixed position. Therefore, the data of devices can be efficiently collected by optimizing the hovering location and moving path of UAVs. 

\par However, deploying a UAV in such a scenario also confronts with several challenges. First, when a UAV is deployed in one location, it often covers multiple sensor devices. The devices that are far away from the UAV will experience more complex and longer transmission channels, resulting in relatively lower transmit rate. Second, different distances between the devices and UAV will bring uneven energy consumption for data transmission, which indicates that the devices with longer distance will consume more energy for transmitting data to UAV. Finally, during the deployment of a UAV, it will consume corresponding energy of a UAV, which will affect the communication performance and lifetime of the system.

\par To sum up, there is a balanced relationship between the challenges abovementioned, which must be comprehensively taken into account. In addition, the speed of of a UAV will have an effect on the moving energy consumption of UAV, therefore it needs to be considered. Moreover, the transmit power of the devices can also affect the lifetime of the system. Therefore, we formulate a UDCMOP, to simultaneously maximize the minimum transmit rate of devices, minimize the total energy consumption of devices, and minimize the total energy consumption of a UAV by cooperatively optimizing the deployment of a UAV $[\mathbb{X}, \mathbb{Y}] = \{{x}_1, ..., x_u, ..., x_U, {y}_1, ..., y_u, ..., y_U\}$, visit sequence of a UAV at different locations $\mathbb{C} = \{C_1, ..., C_u, ..., C_U\}$, speed of a UAV $\mathbb{V} = \{V_1, ..., V_u, ..., V_U\}$, and transmit power of devices $\mathbb{P}=\{p_1, ..., p_K\}$. Consequently, the solution of the formulated UDCMOP can be stated as $X = [\mathbb{X}, \mathbb{Y}, \mathbb{C}, \mathbb{V}, \mathbb{P}]$, followed by a description of the corresponding optimization goals.

\textbf{\emph{1) Objective 1: Maximize the minimum transmit rate of devices (MMTRD)}}. Data transmit rate can well evaluate the performance of data transmission of devices in UAV-assisted agricultural IoT system. In order to increase the fairness of system, we take into account maximizing the minimum transmit rate of all devices, and the first objective function can be expressed as follows:

\begin{equation}
	\begin{aligned}
		&f_1(X) = \min _{k} \{R_{k, u}\}, k \in K,
	\end{aligned}
\end{equation}

\textbf{\emph{2) Objective 2: Minimize the total energy consumption of devices (MTECD)}}. One of the most key indicators to evaluate the effectiveness of the UAV-assisted agricultural IoT data gathering system is the energy consumption of the devices. Therefore, we consider minimizing the total energy consumed by all devices, and the second objective function is written as follows:

\begin{equation}
	\begin{aligned}
		&f_2(X) = \sum_{k=1}^{K} E_{k, u},
	\end{aligned}
\end{equation}

\textbf{\emph{3) Objective 3: Minimize the total energy consumption of a UAV (MTECU)}}. In order to collect data under the consideration of the above two conditions, UAV needs to find the optimal hovering position, which will increase the corresponding mobile energy consumption. In addition, when devices upload data to the UAV at the corresponding location, the UAV remains in a hovering state, which will generate hovering energy consumption. Therefore, we define the third objective function as follows:

\begin{equation}
	\begin{aligned}
		&f_3(X) = {E^{{u}} + E^{m}},
	\end{aligned}
\end{equation}

\par According to the descriptions above, the UDCMOP in the considered UAV-assisted data collection agriculture IoT can be formulated as follows:

\begin{subequations}
	\begin{align}
		\min _{X} & F=\left\{-f_{1}, f_{2}, f_{3}\right\} \\
		\text { s.t. } 
		& C 1: X_{\min } \leqslant x_{u} \leqslant X_{\max }, \forall u, \\
		& C 2: Y_{\min } \leqslant y_{u} \leqslant Y_{\max }, \forall u, \\
		& C 3: P_{\min} \leqslant p_{k} \leqslant P_{\max}, \forall k,\\
		& C 4: V_{\min} \leqslant V_{u} \leqslant V_{\max}, \forall u, \\
		& C 5: c_{k,u} \in \{0, 1\}, \forall k, \forall u, \\
		& C 6: \mathbb{C} \in \{C_1, ..., C_U\},
	\end{align}
	\label{UDCMOP}
\end{subequations}

\noindent where constraints $C1$ and $C2$ together specify the movement area of a UAV. Constraint $C3$ ensures that each device transmits data with the power within the power constraints. $C4$ limits the speed of a UAV in various path segments. Moreover, $c_{k,u}$ is used to represent the association relationship between the devices and UAV. The constraint $C6$ is used to limit the visit sequence of a UAV, in which $U$ is the number of UAV hovering, and it is also the amount of subareas. For ease of searching, we generalize the notations used throughout the paper in Table \ref{Summary of notations.}.

\begin{table}[htb]
	\renewcommand\arraystretch{1.2}
	\centering
	\caption{Summary of notations.}
	\resizebox{1\columnwidth}{!}{
		\begin{tabular}{|c|c|} \hline
			{\bfseries Notation} & {{\bfseries Physical meaning}} \\\hline
			B & Bandwidth \\\hline
			W & Aircraft weight in kg \\\hline	
			$\rho$ & Air density in kg/m${^3}$ \\\hline	
			R & Rotor radius in meter (m) \\\hline
			$A_a$ & Rotor disc area in m${^2}$, $A_a$ $\triangleq$ $\pi$R$^2$ \\\hline
			$\Omega$ & Blade angular velocity in radians/second \\\hline
			$U_{tip}$ & Tip speed of the rotor blade (m/s), $U_{tip}$ $\triangleq$ $\Omega$R \\\hline
			s &Rotor solidity \\\hline
			$d_0$ &Fuselage drag ratio \\\hline
			$v_0$ & Mean rotor induced velocity in hover \\\hline
			$p_0$ & Blade profile power in hovering status \\\hline 
			$V$ & The movement speed of UAV \\\hline
			$P_i$ & Induced power in hovering status \\\hline
			$\mu_0$ & Additional signal attenuation factor due to the NLoS propagation \\\hline
			$\alpha_L$ &Average path loss exponents for the LoS \\\hline
			$\alpha_N$ &Average path loss exponents for the NLoS \\\hline
			$C$ & Constant depending on the environment \\\hline
			$B_c$ & Constant depending on the environment \\\hline
			$\sigma ^2$ & Noise variance \\\hline
			$\beta_0$ & Average reference channel gain \\\hline
	\end{tabular}}
	\label{Summary of notations.}
\end{table}

\subsection{Problem analysis}
\subsubsection{Trade-offs} 
\par In the formulated UDCMOP, if the device transmits data with a smaller minimum transmit rate, it will consume less energy of evices. However, it will result in a longer hovering time for UAV, and consume higher hovering energy of UAV. The energy consumption of hovering and moving of UAV are composed of total energy consumption of UAV, therefore the total energy consumption will also increase. On the contrary, if the devices transmit data at a higher rate, it will consume a larger amount of transmit power of devices. Then the hovering time of UAV is shortened, and the consumed hovering energy is reduced, leading to smaller total energy consumption of the UAV. In summary, there is a balanced relationship between the three objectives abovementioned.

\par On the other hand, problem (\ref{UDCMOP}) essentially involves simultaneously optimizing the three objective functions (i.e., $f_ 1$, $f_ 2$, and $f_3$) via jointly optimizing $\mathbb{X}, \mathbb{Y}, \mathbb{C}, \mathbb{V}$, and $\mathbb{P}$. However, with respect to the hovering position $\{\mathbb{X}, \mathbb{Y}\}$ and visit sequence $\mathbb{C}$ of UAV, most of the existing works are to separately optimize them \cite{wu2018joint}, \cite{samir2019uav}, while the problem of simultaneously optimizing hovering point and visit sequence of UAV is currently not solved. Therefore, in this work, we have carried out a series of works to address this issue.

\par The developed UDCMOP is not a continuous decision problem, but a discrete and continuous combination decision problem. In addition, there is a balanced relationship between these objectives, and it is complicated to be solved, while MOEAs can be used to solve such problems. Therefore, we employ one of MOEAs, namely MOAHA, as the basic algorithm to solve our developed UDCMOP, and the details are provided in Section \ref{Methodology}.

\subsubsection{NP-hardness}
\par In proposed UDCMOP, the third part of the solutions, namely the visit sequence $\mathbb{C}$ of UAV, mainly has an effect on the moving distance of the UAV, which ultimately has an impact on the energy consumed by UAV, namely the third objective of problem (\ref{UDCMOP}). Moreover, we consider a special case of problem (\ref{UDCMOP}), where the UAV starts from the starting position to the end position, hovering once over each subarea, and there is no intersection on the path to the end position. The determined visit sequence can make the path of the UAV shortest, so that minimizing the energy consumption of the UAV. This problem is essentially the same as the selective traveling salesman problem (TSP) (or orienteering problem), known as NP-hard \cite{vansteenwegen2011orienteering}, \cite{zhan2018trajectory}.

\par Therefore, problem (\ref{UDCMOP}) is also NP-hard when the UAV hovers over other subareas, and it is more challenging to acquire the optimal solution than TSP.

%
%
\section{Methodology}
\label{Methodology}

\par In this section, an IMOAHA method is suggested to tackle the formulated UDCMOP of UAV-assisted agricultural IoT.

%
%
\subsection{Motivation}
\par The three optimization objective functions in formulated UDCMOP need to be optimized at the same time. Furthermore, the developed UDCMOP is a nonlinear mixed integer programming problem, causing it more complicated and difficult to be solved via utilizing traditional mathematical methods, such as convex/concave optimization, in addition to the gradient descent method. However, as one of the popular methods for solving multiple objective functions simultaneously, MOEAs typically combine these functions and there is no requirement to separate them into multiple independent single objectives or linearly weight these objective functions to solve.

\par Moreover, compared to other methods, the MOEAs require few parameters to be adjusted, and they have the fast convergence rate. Specifically, they can find the optimal solution in a reasonable calculation time, and can effectively solve the NP-hard problem. In addition, they have attracted many attention of researchers because of their ability to solve multidimensional, multimodal, combinatorial and large-scale search space problems, such as, multi-objective whale optimization algorithm (MOWOA) \cite{kumawat2017multi}, multi-objective atomic orbital search (MOAOS) \cite{azizi2022multiobjective}, in addition to the multi-objective crystal structure algorithm (MOCryStAl) \cite{khodadadi2021multi}.

\par The AHA is a novel population based algorithm \cite{ramadan2022accurate}, and is inspired by the distinct flight ability and foraging behavior of hummingbirds. The distinct foraging, flight, and unique memory updating mechanisms of hummingbirds cooperatively make AHA vastly different from other algorithms \cite{bhagat2023application}. Moreover, AHA has the capacity and competitiveness in terms of computational accuracy and time, and can rapidly and accurately discover the global optimal solution comparing to other algorithms \cite{chen2023pso}. In addition, due to its superior performance and characteristics, AHA is a potential candidate for conversion to a multi-objective version for the purpose of solving MOPs \cite{zhao2022effective}. Therefore, the multi-objective AHA (MOAHA) \cite{zhao2022effective} is suggested and applied into actual engineering MOPs, and shows excellent performance. Thereupon, we employ it to solve the formulated UDCMOP. However, this method faces with some issues when dealing with the UDCMOP provided in (\ref{UDCMOP}):

1) The formulated UDCMOP is composed of continuous and discrete solutions, i.e., ($\mathbb{X}, \mathbb{Y}, \mathbb{V}, \mathbb{P}$) and $\mathbb{C}$, which makes it laborious to be solved by using the conventional MOAHA. 

2) In order to solve problem (\ref{UDCMOP}), we collaboratively optimize the horizontal location of UAV ($\mathbb{X}, \mathbb{Y}$), the visit sequence of UAV $\mathbb{C}$, the speed of UAV $\mathbb{V}$, and the transmit power of devices $\mathbb{P}$. Therefore, the solution dimensions can be denoted as (4 $\times$ $U$ + $K$). It can be seen that the scale of the solution increases as the increase of hovering times of UAV and amount of IoT devices, which means that the increasing of $U$ and $K$ will have a great effect on the scale of the formulated UDCMOP.

3) The physical meaning of each component of solutions in the UDCMOP is different, which means that their value ranges are also different. Moreover, the standard MOAHA can be used to solve the problems of continuous solutions, while they are not suitable for solving the problems of hybrid solutions.

\par Therefore, all of the aforementioned difficulties have driven us to enhance the standard MOAHA in an effort to deal with the developed UDCMOP.

%
%
\subsection{Conventional MOAHA}

\par The foraging and flight behavior of hummingbirds in nature, particularly their guided, territorial, and migratory foraging behaviors, served as an inspiration for the invention of AHA. In addition, three flight strategies are considered in the process of foraging, e.g., the axial flight, diagonal flight, and omnidirectional flight. MOAHA mainly includes three mechanism to make it suitable for solving MOPs, that are an external archive mechanism, a dynamic elimination-based crowding distance (DECD) method, and a solution update mechanism based on non-dominated solutions (NDS). 

\subsubsection{External archive}
\par In order to preserve a fixed number of optimal non-dominated solutions during the iteration process, an external archive is introduced into MOAHA. In the initial population, all non dominated solutions will be saved in the archive.

\subsubsection{Solution update mechanism based on NDS}
\par In NDS strategy, all solutions are sorted according to NDS, with solutions dominated by fewer solutions having higher dominance levels, and all solutions are ranked according to their non-dominated levels. Three update expression are as follows:

\begin{equation}
	x_i(t+1) = \left\{\begin{array}{ll}
		v_{i}(t+1), ~~v_i \in F_p, x_i \in F_q, ~\rm{and} ~p<q \\
		\begin{cases}
			v_{i}(t+1), v_i \in F_p, x_i \in F_q, p=q, \rm{and} ~rand <0.5, \\
			x_{i}(t), ~~ \quad v_i \in F_p, v_i \in F_q, p=q, \rm{and} ~rand \geq 0.5, \\
		\end{cases} \\
		x_i(t), \quad ~~~~v_i \in F_p, x_i \in F_q ~\rm{and}~ p>q, \\
	\end{array}\right.
	\label{universe population}
\end{equation}

\noindent where $F_p$ and $F_q$ are the $p$th and $q$th fronts in NDS. $v_{i}(t+1)$ is the candidate solution at the $(t+1)$th iteration. $x_i(t)$ and $x_i(t+1)$ are the positions of the $i$th food source at iterations $t$ and $t+1$.

\subsubsection{Dynamic elimination-based crowding distance(DECD)}
\par To increase the variety of solutions, the crowding distance is introduced as a powerful parameter-free technique \cite{deb2002fast}, which can maintain a fixed-size external archive after removing excessive solutions with minimum crowding distance~\cite{zhao2022artificial}. Whenever a solution is removed, the crowding distance ranking of the remaining solutions will change, reducing the diversity of solutions. This means that the archiving process based on crowding distance is not conducive to maintaining the sustainable diversity of Pareto set (PS). Therefore, an external archive with DECD is proposed. In PS, after the solution with the minimum crowding distance is removed, only the crowding distance of the solution closest to the removed solution is updated, and the crowding distance of the remaining solutions remains unchanged. Moreover, the pseudocode and details of MOAHA can be found in \cite{zhao2022effective}.

%
%
\subsection{IMOAHA}

\par For the purpose of solving the defined UDCMOP, we suggest an IMOAHA in this section. The performance of the IMOAHA is strengthened by incorporating three improved factors, that are the hybrid initialization operator, Cauchy mutation foraging operator, as wel as the discrete mutation operator, and the details are stated as follows.

%
%
\subsubsection{Hybrid initialization operator}

\par In MOEAs, including MOAHA, the population is randomly initialized, which makes the solution easy to sink into the local optimum during the evolution process. Therefore, it is significant to strengthen the initialization operator of the solutions.

\par Chaotic systems are characterized by ergodicity, periodic oscillation, sensitivity to initial conditions, and inherent randomness \cite{yuan2008hydrothermal}, making them widely applied in various applications, including engineering fields. Moreover, it is shown in many numerical results that the initialization of algorithm in search space can be improved by using chaotic sequences in optimization algorithm \cite{yuan2008hydrothermal}. The sensitivity of chaotic sequences to initial conditions makes them show unpredictable long-term behavior, which helps to track chaotic variables \cite{yuan2008hydrothermal}. Therefore, in order to strengthen the quality of the solutions, a hybrid initialization operator is proposed, which consists of chaotic sequences and random sequences. The former is used to generate a set of continuous variables of MOAHA, and the latter is used to initialize discrete variables.

\par In this paper, the Tent map is used to generate continuous variables of UDCMOP (including the horizontal coordinates of UAV, the speed of UAV, in addition to the transmit power of devices), and the expression is designed as follows:

\begin{equation}
	\label{Tent map}
	rand_{x1}= 
	\begin{cases}
		rand_{x0}/d \cdot rand, & {\rm if} \quad rand_{x0} < 0.7, \\
		e \cdot (rand_{x0} - 1) \cdot rand, &{\rm otherwise}, \\
	\end{cases}
\end{equation}

\begin{equation}
	x = L +	rand_{x1} \cdot (U - L),
	\label{continuous variables}
\end{equation}

\noindent where $d$, $e$, and $rand_{x0}$ are three constants \cite{saremi2014biogeography}. $rand$ is a random number between (0, 1). $rand_{x1}$ denotes the random number of continuous variables used to generate UDCMOP solutions. In addition, $L$ and $U$ are the minimum and maximum values of continuous variables, respectively. As mentioned above, in the developed UDCMOP, the dimensions of the optimization variables are composed of a variety of diverse physical meanings, which indicates that the associated search ranges are also varied \cite{liu2022multiobjective}. Therefore, during initialization, in terms of the solutions with distinct physical meanings, we should set distinct value ranges, which means that the values of $L$ and $U$ are various with respect to diverse physical meanings. In addition, for the discrete part of the solutions, we initialize it according to the ascending order of the number of UAV hovering points.

%
%
\subsubsection{Cauchy mutation foraging operator}

\par The Cauchy mutation component is added into the global search phase to enhance the capacity of global search of IMOAHA. The random mutation process known as ``Cauchy mutation" is based on the Cauchy distribution, and has been shown to have strong convergence and the ability to avoid local optimum \cite{ali2011improving}. Moreover, Cauchy mutation has been combined with meta-heuristic algorithms to strengthen the performance of original methods, like differential evolutionary (DE) \cite{ali2011improving}, krill herd (KH) \cite{wang2016opposition}. The probability density function of Cauchy mutation is provided as follows \cite{wang2016opposition}:

\begin{equation}
	F(x) = \frac{1}{2} + \frac{1}{\pi}\arctan(\frac{x}{t}),
\end{equation}

\noindent where $t > 0$ represents the scalar factor. Moreover, in original MOAHA, the proportion of guided foraging is equal to that of territorial foraging, and a hummingbird can either perform guided foraging or territorial foraging, which will result in a smaller search scope and poor diversity of solutions. Therefore, Cauchy mutation is used to conduct another mutation on the basis of guided foraging, which improves the exploration ability of the solutions, and enhances the evolutionary performance of the population \cite{zhao2021enhanced}. Note that Cauchy mutation is only applied to the evolution of horizontal position and speed of UAV. The specific expression of Cauchy mutation foraging operator is calculated as follows:

\begin{equation}
	\begin{aligned}
		Pop_{new} &= Arch_{pop}(randi([1 \quad q])) + Arch_{pop}(randi([1 \quad q])) \\
		& \cdot E \cdot 1/(\pi \cdot(rand^2 +1)), 
	\end{aligned}
	\label{Cauchy mutation}
\end{equation}

\noindent where $Pop_{new}$ denotes the new individual generated by Cauchy mutation. $Arch_{pop}$ represents the optimal individual stored in archive, and $randi()$ is used to generate a random integer number between 1 and $q$, wherein $q$ is the length of respective individual with various physical meanings. Note that the length of each solution component of the formulated UDCMOP is different, which means that the value of $q$ is different with respect to the horizontal position and speed of UAV. Moreover, $E = 0.1 \cdot rand$ and 0.01 for the horizontal position and speed of UAV, respectively, which are determined by their respective search scope. The Cauchy mutation foraging operator is given in Algorithm \ref{algorithm: 1}.

\begin{algorithm}
	\caption{Cauchy mutation foraging operator.}
	\label{algorithm: 1}
	\begin{algorithmic}[1]
		\Require $nPop$ denotes the population size; $Iter_{max}$ represents the maximum iteration numbers; $L$ and $U$ are the range of variables; $Arch_{max}$ constraints the size of archive; \\
		Perform discrete mutation operator according to Algorithm \ref{algorithm:2};
		\If {$rand <$ 0.5} \\
		\quad Perform guided foraging;
		\If {$rand <$ 0.2} \\
		\quad \quad Refer to Eq. (\ref{Cauchy mutation}), Cauchy mutation is performed on the corresponding dimensions of individual in the population.
		\EndIf 
		\Else \\
		\quad Perform territorial foraging;
		\EndIf 
	\end{algorithmic}
\end{algorithm}

\subsubsection{Discrete mutation operator}

\par In this section, we propose a discrete mutation operator to optimize the discrete part of the solution, in order to compensate for the shortcomings of conventional MOEAs that only solve continuous solutions.

\begin{figure}[htbp]
	\centering
	\includegraphics[width=3.4in]{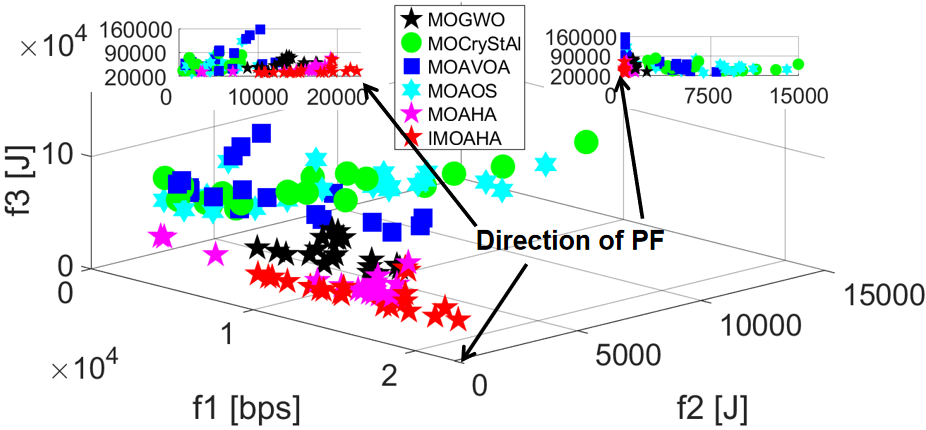}
	\caption{Solution distributions obtained by different algorithms in Case 1 (6 UAV hovering points).}
	\label{solution distributions6}
\end{figure}

\par In the phase of guided foraging, hummingbirds will visit the food source that has the largest quantity of nectar-refilling. The food source with the largest amount of nectar-refilling represents that the hummingbirds have the optimal objective function values. Therefore, in guided foraging stage, the visiting order of hummingbirds will inherit the visiting order of the first individual in population, which ensures that the foraging behavior of hummingbirds can be updated with the renewal of population, and enhance the exploration capacity of the solutions. In addition, the optimal non-dominated solution in MOAHA will be saved to the external archive. Therefore, in territory foraging phase, we randomly select an individual among the external archive to enable the hummingbirds to forage according to the visit order of the individuals, which can ensure the hummingbirds to update among the optimal solutions and speed up the convergence. It is worth pointing out that in the process of updating solutions of the improved algorithm and other comparison algorithms, the discrete parts are updated by using discrete mutation operator, and the specifics are shown in Algorithm \ref{algorithm:2}.

\begin{algorithm}
	\caption{Discrete mutation operator.}
	\label{algorithm:2}
	\begin{algorithmic}
		\Require Refer to Algorithm \ref{algorithm: 1}; \\
		\textbf{Discrete mutation operator}: \\
		Generate: Randomly generate a set of visit sequences of UAV $\mathbb{C}$; \\
		Select: Randomly select two individuals $index1$ and $index2$ from $\mathbb{C}$; \\
		Judge and exchange: Judge whether the values of these two individuals are equal, and if not, exchange them; 
		\Ensure Output a new sequence.
	\end{algorithmic}
\end{algorithm}

\renewcommand{\dblfloatpagefraction}{.9}
\begin{figure}[htbp]
	\centering
	\includegraphics[width=3.4in]{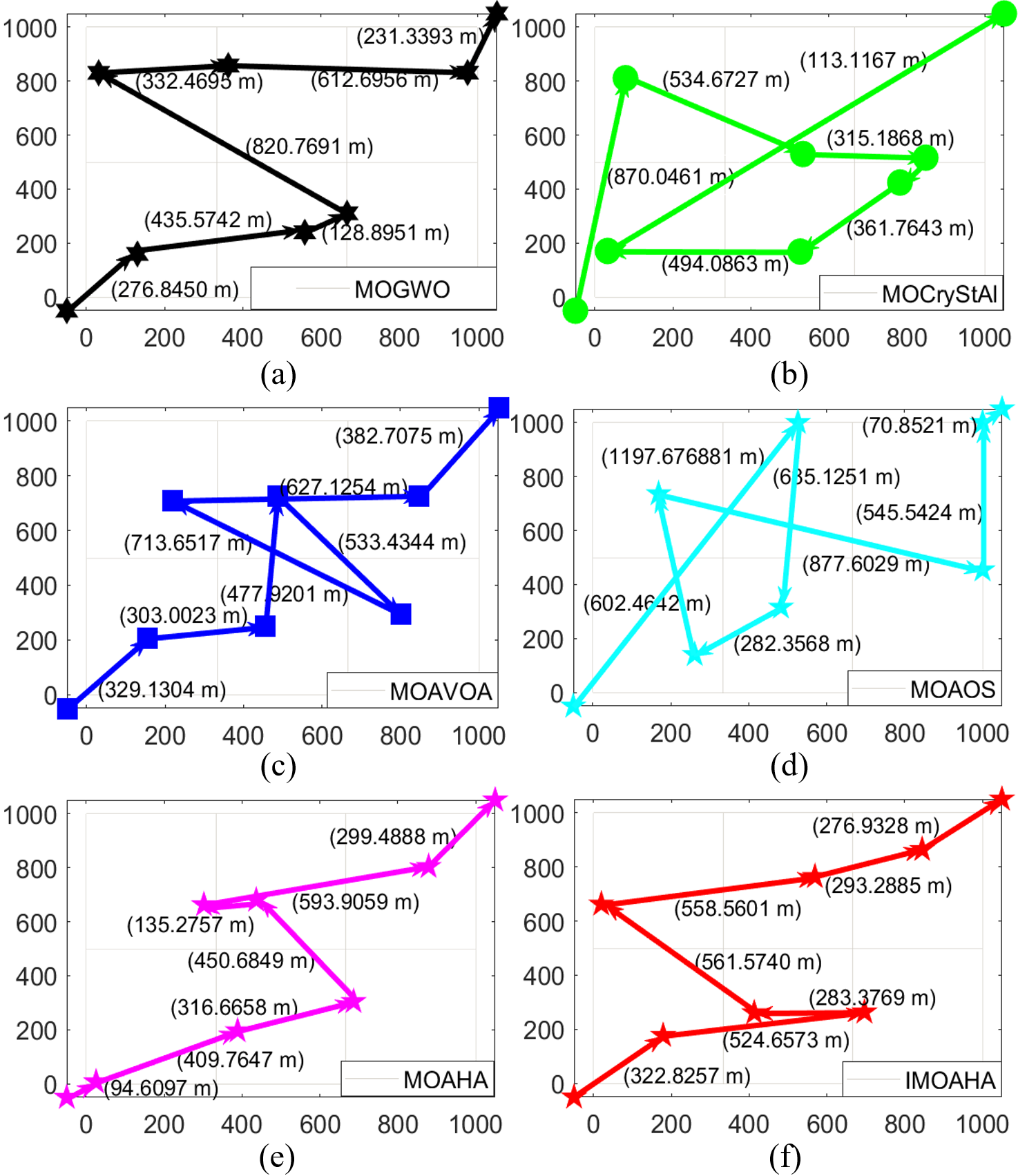}
	\caption{Movement path of 6 UAV hovering points obtained by different algorithms.}
	\label{Movement path of 6 UAV hovering points obtained by different algorithms.}
\end{figure}

%
%
\subsection{Complexity of the proposed algorithm}
\par The computational complexity $O$(IMOAHA) of IMOAHA is mainly calculated as $O$(non-dominated sorting) + $O$(Deleting element and updating archive) + $O$(Calculating crowding distance)+ $O$(dynamic elimination-based crowding distance) + $O$(Performing the foraging phase). These parameters will be determined by the population size $N$, the maximum iteration numbers $It$, and the number of objective functions $M$. $d$ and $D$ are the dimensions of discrete and continuous solution in solution space, respectively. In this work, to facilitate analysis, the size of the external archive is set to be the same as that of the population. 

\par Although the hybrid initialization operator enhances the quality of initialization solutions of IMOAHA, the computational complexity remains unchanged comparing to conventional MOAHA. In foraging phases, the complexity of discrete mutation operator is $O(TNd)$. The computational complexity of the original guided foraging is $O(0.5TMN)$, while the Cauchy mutation foraging operator has $O(0.2TMN)$ complexity. The non-dominated sorting requires $O(TMN^2)$ computational complexity.

\par When updating the archive and removing elements with the least crowding distance from population, the computational difficulty is $O(TMN^2 logN)$. Calculating the crowding distance between two adjacent elements has a computational complexity of $O(2TMN)$. Therefore, it has $O(TMN^2 logN)$ computational complexity in dynamic elimination-based crowding distance operator. Moreover, in territorial and migration foraging phases, $O(TMN)$ and $O(0.5TM/N)$ computational complexities are required, respectively. To sum up, the overall complexity of IMOAHA is $O(TMN^2 logN)$.

%
%
\section{Simulation results and analysis}
\label{Simulation results and analysis}
\subsection{Simulation setups}

\par In this work, we conduct a series of simulations in Matlab 2021a for verifying the performance of IMOAHA to solve the formulated UDCMOP. The computer used for the simulation is with a AMD (R) 7 5700G Radeon Graphics 3.8 GHz, and the RAM is 16 G.

\par It is assumed that the size of the agriculture area is set to 1 km $\times$ 1 km, and the altitude of UAV is set to 100 m. Then the minimum and maximum power of UAV (i.e., $P_{min}$ and $P_{max}$) are respectively set to 0.1 W and 10 W. The aircraft weight of UAV $W$ is 2 kg. $\rho$ and $R$ are set to 1.225 kg/m${^3}$ and 0.4 m, respectively. The rotor disc area $A_a$, tip speed of the rotor blade $U_{tip}$, and blade angular velocity $\Omega$ are set to 0.503 m${^2}$, 120 m/s and 300 r/s, respectively. $s$ and $d_0$ are set to 0.05 and 0.6, respectively. $v_0$ and $p_0$ in hovering are set to 4.03 and 79.8563, respectively. Moreover, $P_i$ is set to 96.6850 w. The additional signal attenuation factor due to NLoS propagation is -20 dB. For LoS and NLoS, the average path loss exponents are respectively 2.5 and 3.5. The two constants depending on environment $C$ and $B_c$ are set to 11.95 and 0.136, respectively. The channel bandwidth and noise variance are given by $B$ = 10 MHz and $\sigma^2$ = -110 dBm, respectively. The reference channel gain $\beta_0$ is set to -60 dB. The speed $V$ of UAV in different path segments can be randomly sampled from (10, 20) in m/s. $d=0.7$, $e=10/3$ and $rand_{x0} = 0.6$. $t=1$ \cite{wang2016opposition}. In addition, for the purpose of avoiding random bias, the maximum iteration number is set to 200, and each algorithm is run 30 times. The remaining parameters can refer to \cite{zeng2019energy}.

\par In this paper, we organize two groups of comparative experiments to verify the performance and effectiveness of the IMOAHA. On the one hand, in terms of the hovering number of UAV, we set it to 6 and 8, which correspond to the number of farmland blocks in the entire agricultural area. On the other hand, for the comparative algorithms, several comparison methods, such as MOCryStAl, multi-objective artificial vultures optimization algorithm (MOAVOA) \cite{khodadadi2022moavoa}, MOAOS, the conventional MOAHA, and the proposed IMOAHA are introduced to testify the performance of the proposed algorithm. 

%
%
\subsection{Optimization results}
\par In this section, the optimization outcomes obtained by the proposed IMOAHA and other comparison methods are presented. 
	\begin{figure}[htbp]
		\centering
		\includegraphics[width=3.4in]{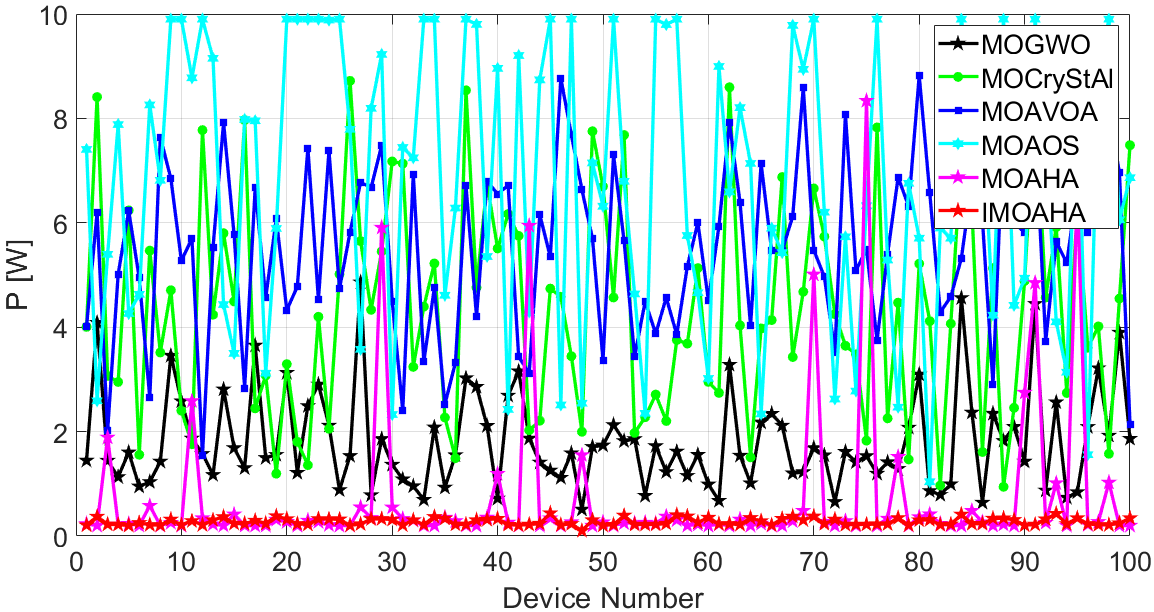}
		\caption {Transmission power consumed by all devices under different algorithms in Case 1 (6 UAV hovering points).}
		\label{Transmission power6}
	\end{figure}

%
%
\subsubsection{Case 1: The optimization results of 6 UAV hovering points}

\par In this section, in an effort to verify the performance and effectiveness of the suggested IMOAHA, UAV communicates with 100 sensor devices in the farmland by hovering in 6 optimal positions. The numerical results achieved by different methods mentioned above are given in Table \ref{Numerical results obtained by different methods for case 1 (6 UAV hovering points)} in terms of MMTRD $f_1$, MTECD $f_2$, as well as MTECU $f_3$. As can be seen, the suggested IMOAHA performs best on all optimization objectives.

	\begin{table}[htbp]
		\centering
		\caption{Numerical results achieved by different methods for case 1 (6 hovering points).}
		\begin{tabular}{|c|c|c|c|}
			\hline
			{\bfseries Algorithm} & {$f_1$({\bfseries bps})} & {$f_2$({\bfseries J})} & {$f_3$({\bfseries J})} \\\hline
			MOGWO & 13626.3592 & 1014.0016 & 58698.0896 \\\hline
			MOCrystal & 3898.8019 & 6003.8599 & 54867.0351 \\\hline
			MOAVOA & 6062.8155 & 3252.3388 & 71387.6427 \\\hline
			MOAOS & 4585.2640 & 6935.7325 & 52027.5845 \\\hline
			MOAHA & 15048.1190 & 403.3498 & 42636.2888 \\\hline
			$\textbf{IMOAHA}$ & $\bm{16435.3817}$ & $\bm{135.8534}$ & $\bm{38254.0313}$ \\\hline
		\end{tabular}
		\label{Numerical results obtained by different methods for case 1 (6 UAV hovering points)}
	\end{table}

	\begin{table}[htbp]
		\centering
		\caption{The trajectory length of different algorithms of two cases.}
		\begin{tabular}{|c|c|c|} \hline
			\thead{\bfseries Algorithm} & \thead{\bfseries UAV trajectory length \\ \bfseries of Case 1 ({\bfseries m})} & \thead{\bfseries UAV trajectory length \\ \bfseries of Case 2 ({\bfseries m})} \\\hline
			MOGWO & 2838.5878 & 4257.6000 \\\hline
			MOCrystal & 4033.4117 & 4587.1870 \\\hline
			MOAVOA & 3366.9719 & 4806.1397 \\\hline
			MOAOS & 4261.6204 & 4915.4896 \\\hline
			MOAHA & 2300.3954 & 3095.7916 \\\hline
			IMOAHA & 2821.2000 & 3967.8072 \\\hline
		\end{tabular}%
		\label{The trajectory length of different algorithms}%
	\end{table}
	
	\begin{figure}[htbp]
		\centering
		\includegraphics[width=3.4in]{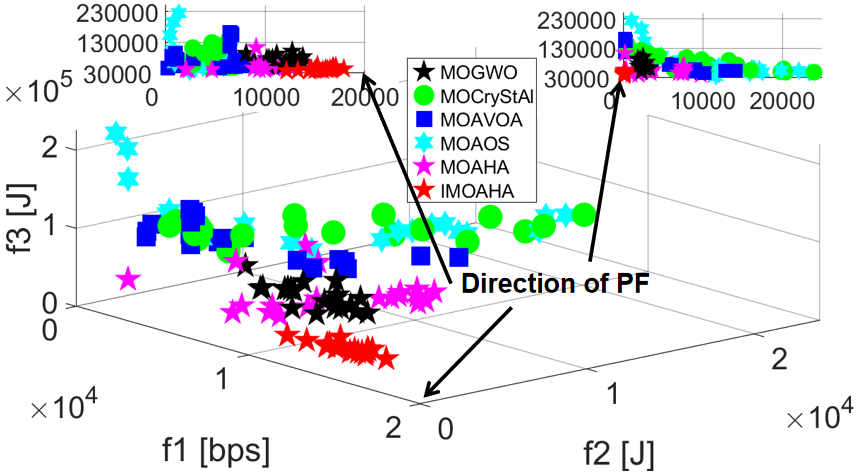}
		\caption{Solution distributions obtained by different algorithms in Case 2 (8 UAV hovering points).}
		\label{solution distributions8}
	\end{figure}
	
	\begin{figure}[htbp]
		\centering
		\includegraphics[width=3.4in]{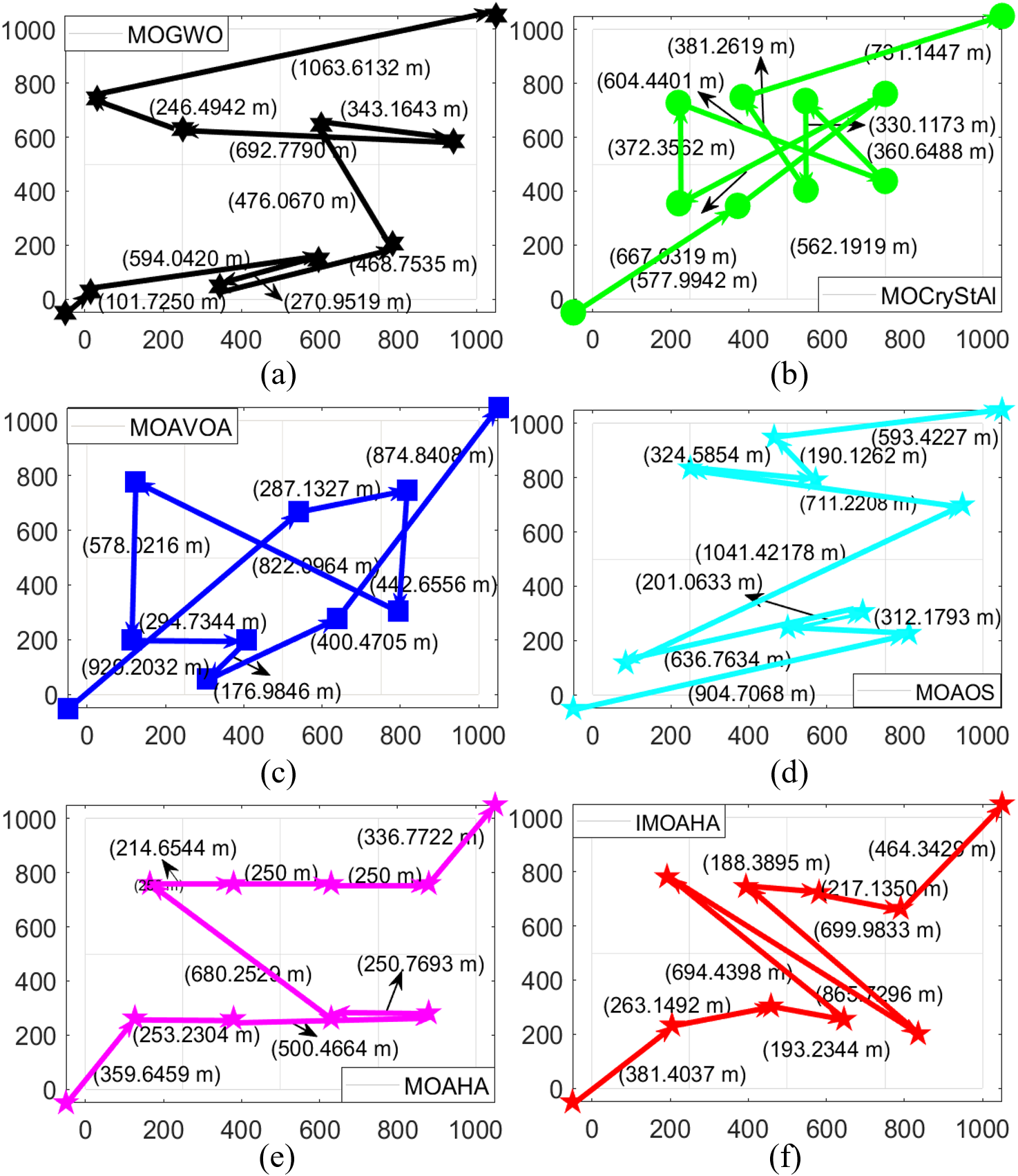}
		\caption {Movement path of 8 UAV hovering points obtained by different algorithms.}
		\label{Movement path of 8 UAV hovering points obtained by different algorithms.}
	\end{figure}
\par In addition, the distribution results of solutions obtained by different MOEAs are shown in Fig. \ref{solution distributions6}. It can be seen that the suggested IMOAHA approach is more appropriate for solving the defined UDCMOP than other approaches since the solutions it produces are closer to the optimal solution (i.e., the PF depicted in these figures) than those produced by other methods. It is noted that the values listed here are the average of 200 iterations.

\par The movement path and corresponding trajectory length of UAV in case 1 during communicating with devices are shown in Fig. \ref{Movement path of 6 UAV hovering points obtained by different algorithms.} and Table \ref{The trajectory length of different algorithms}, which mean that the UAV hovering at each position can communicate with the devices in the monitoring area. Moreover, the transmit power consumed by all devices under diverse algorithms is shown in Fig. \ref{Transmission power6}.

\subsubsection{Case 2: The optimization results of 8 UAV hovering points}

\par In this section, UAV communicates with devices in the farmland by hovering in 8 optimal positions for verifying the performance and effectiveness of the suggested IMOAHA. The statistical results with respect to three optimization objectives are summarized in Table \ref{Numerical results obtained by different methods for case 2 (8 UAV hovering points}, from which we can learn that the suggested IMOAHA has the highest performance on all objectives.

	\begin{table}[htbp]
		\centering
		\caption{Numerical results achieved by different methods for case 2 (8 hovering points).}
		\begin{tabular}{|c|c|c|c|} \hline
			{\bfseries Algorithm} & {$f_1$({\bfseries bps})} & {$f_2$({\bfseries J})} & {$f_3$({\bfseries J})} \\\hline
			MOGWO & 11512.1090 & 2403.7444 & 74289.4416 \\\hline
			MOCryStAl & 4959.8427 & 9925.9969 & 86265.9054 \\\hline
			MOAVOA & 4753.8859 & 5969.7153 & 84679.0430 \\\hline
			MOAOS & 3690.1924 & 12000.1396 & 83208.7311 \\\hline
			MOAHA & 9743.3783 & 5148.5064 & 54138.2853 \\\hline
			$\textbf{IMOAHA}$ & $\bm{15518.6787}$ & $\bm{156.9246}$ & $\bm{43270.9266}$ \\\hline
		\end{tabular}
		\label{Numerical results obtained by different methods for case 2 (8 UAV hovering points}
	\end{table}
	
	\begin{figure}[htbp]
		\centering
		\includegraphics[width=3.4in]{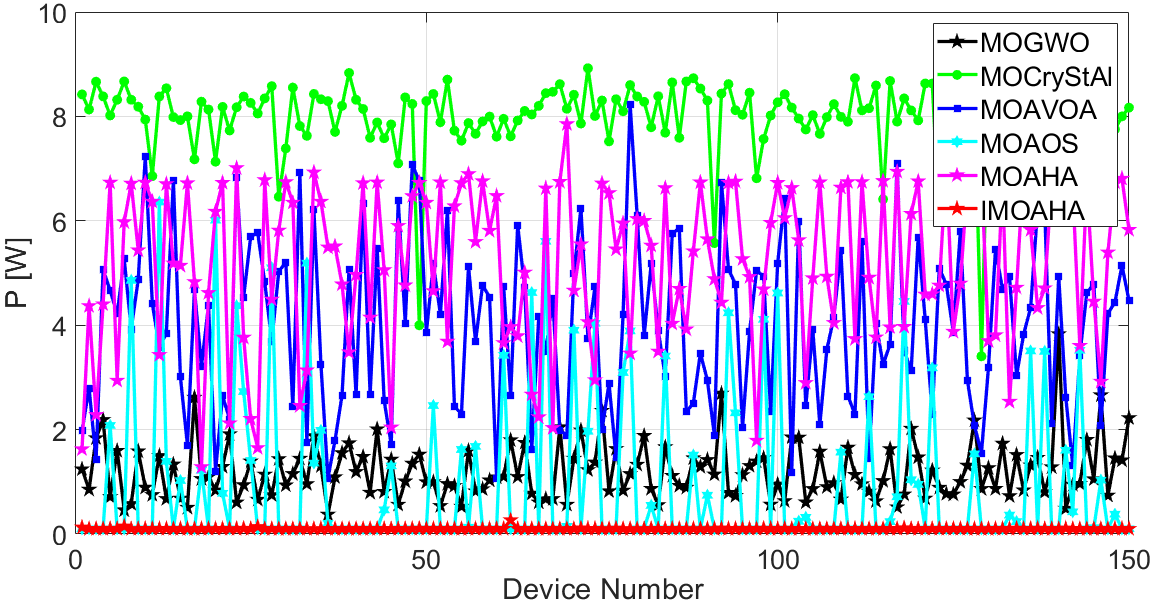}
		\caption {Transmission power consumed by all devices under different algorithms in Case 2 (8 UAV hovering points).}
		\label{Transmission power8}
	\end{figure}

\par Similarly, the solution distribution outcomes from various approaches are displayed in Fig. \ref{solution distributions8}. It can be seen from these figures that, the suggested IMOAHA obtains the greater performance since it is closer to PF. The movement path and trajectory length of UAV in case 2 when it communicates with farmland devices in different subarea are given in Fig. \ref{Movement path of 8 UAV hovering points obtained by different algorithms.} and Table \ref{The trajectory length of different algorithms}, respectively. Moreover, Fig. \ref{Transmission power8} provides the transmit power of UAV in different trajectory segments obtained by different approaches.

%
%
\section{Conclusion}
\label{Conclusion}

\par In this work, we investigate the UAV-assisted data collection in agricultural IoTs, wherein a UAV is sent out to gather data from IoT devices deployed in agriculture by planing its trajectory. First, we formulate the UDCMOP to simultaneously maximize the minimum transmit rate of devices, minimize the total energy consumption of devices, and minimize the total energy consumption of UAV. In an effort to solve this problem, an IMOAHA with three improved strategies is proposed, in which the hybrid initialization operator contributes to increase the quality of initialization solutions, and Cauchy mutation foraging operator is used to improve the ability of global search of IMOAHA, as well as the discrete mutation operator is benefit to optimize discrete solutions. Final, simulation outcomes indicate that with respect to the transmit rate of devices, and the total energy consumption of devices and UAV, our proposed method outperforms other comparative methods, including MOGWO, MOCryStAl, MOAVOA, MOAOS, and conventional MOAHA.

\bibliographystyle{IEEEtran}
\bibliography{refs-UAV-01}

\begin{thebibliography}{10}
\providecommand{\url}[1]{#1}
\csname url@samestyle\endcsname
\providecommand{\newblock}{\relax}
\providecommand{\bibinfo}[2]{#2}
\providecommand{\BIBentrySTDinterwordspacing}{\spaceskip=0pt\relax}
\providecommand{\BIBentryALTinterwordstretchfactor}{4}
\providecommand{\BIBentryALTinterwordspacing}{\spaceskip=\fontdimen2\font plus
\BIBentryALTinterwordstretchfactor\fontdimen3\font minus \fontdimen4\font\relax}
\providecommand{\BIBforeignlanguage}[2]{{%
\expandafter\ifx\csname l@#1\endcsname\relax
\typeout{** WARNING: IEEEtran.bst: No hyphenation pattern has been}%
\typeout{** loaded for the language `#1'. Using the pattern for}%
\typeout{** the default language instead.}%
\else
\language=\csname l@#1\endcsname
\fi
#2}}
\providecommand{\BIBdecl}{\relax}
\BIBdecl

\bibitem{park2021low}
J.~Park, S.~Kim, J.~Youn, S.~Ahn, and S.~Cho, ``Low-complexity data collection scheme for {UAV} sink nodes in cellular iot networks,'' \emph{IEEE Trans. Veh. Technol.}, vol.~70, no.~5, pp. 4865--4879, 2021.

\bibitem{sanchez2014smartsantander}
L.~Sanchez, L.~Mu{\~n}oz, J.~A. Galache, P.~Sotres, J.~R. Santana, V.~Gutierrez, R.~Ramdhany, A.~Gluhak, S.~Krco, E.~Theodoridis \emph{et~al.}, ``Smartsantander: {I}o{T} experimentation over a smart city testbed,'' \emph{Comput. Networks.}, vol.~61, pp. 217--238, 2014.

\bibitem{park2017comprehensive}
E.~Park, Y.~Cho, J.~Han, and S.~J. Kwon, ``Comprehensive approaches to user acceptance of {I}nternet of {T}hings in a smart home environment,'' \emph{IEEE Internet Things J.}, vol.~4, no.~6, pp. 2342--2350, 2017.

\bibitem{zhou2021UAV}
M.~Zhou, H.~Chen, L.~Shu, and Y.~Liu, ``{UAV} assisted sleep scheduling algorithm for energy-efficient data collection in agricultural {I}nternet of {T}hings,'' \emph{IEEE Internet Things J.}, 2021.

\bibitem{tan2019adaptive}
J.~Tan, W.~Liu, T.~Wang, N.~N. Xiong, H.~Song, A.~Liu, and Z.~Zeng, ``An adaptive collection scheme-based matrix completion for data gathering in energy-harvesting wireless sensor networks,'' \emph{IEEE Access}, vol.~7, pp. 6703--6723, 2019.

\bibitem{mostafaei2018energy}
H.~Mostafaei, ``Energy-efficient algorithm for reliable routing of wireless sensor networks,'' \emph{IEEE Trans. Ind. Electron.}, vol.~66, no.~7, pp. 5567--5575, 2018.

\bibitem{zeng2016wireless}
Y.~Zeng, R.~Zhang, and T.~J. Lim, ``Wireless communications with unmanned aerial vehicles: Opportunities and challenges,'' \emph{IEEE Commun. Mag.}, vol.~54, no.~5, pp. 36--42, 2016.

\bibitem{duan2019resource}
R.~Duan, J.~Wang, C.~Jiang, H.~Yao, Y.~Ren, and Y.~Qian, ``Resource allocation for multi-{UAV} aided {I}o{T} {NOMA} uplink transmission systems,'' \emph{IEEE Internet Things J.}, vol.~6, no.~4, pp. 7025--7037, 2019.

\bibitem{ji2019performance}
B.~Ji, Y.~Li, B.~Zhou, C.~Li, K.~Song, and H.~Wen, ``Performance analysis of {{UAV}} relay assisted {I}o{T} communication network enhanced with energy harvesting,'' \emph{IEEE Access}, vol.~7, pp. 38\,738--38\,747, 2019.

\bibitem{aslan2022comprehensive}
M.~F. Aslan, A.~Durdu, K.~Sabanci, E.~Ropelewska, and S.~S. G{\"u}ltekin, ``A comprehensive survey of the recent studies with {UAV} for precision agriculture in open fields and greenhouses,'' \emph{Appl. Sci.}, vol.~12, no.~3, p. 1047, 2022.

\bibitem{lopez2021framework}
A.~L{\'o}pez, J.~M. Jurado, C.~J. Ogayar, and F.~R. Feito, ``A framework for registering {UAV}-based imagery for crop-tracking in precision agriculture,'' \emph{Int. J. Appl. Earth Obs. Geoinformation.}, vol.~97, p. 102274, 2021.

\bibitem{radoglou2020compilation}
P.~Radoglou-Grammatikis, P.~Sarigiannidis, T.~Lagkas, and I.~Moscholios, ``A compilation of {UAV} applications for precision agriculture,'' \emph{Comput. Networks.}, vol. 172, p. 107148, 2020.

\bibitem{popescu2020advanced}
D.~Popescu, F.~Stoican, G.~Stamatescu, L.~Ichim, and C.~Dragana, ``Advanced {UAV}--{WSN} system for intelligent monitoring in precision agriculture,'' \emph{Sensors}, vol.~20, no.~3, p. 817, 2020.

\bibitem{liu2023agricultural}
X.~Liu, G.~Li, H.~Yang, N.~Zhang, L.~Wang, and P.~Shao, ``Agricultural {UAV} trajectory planning by incorporating multi-mechanism improved grey wolf optimization algorithm,'' \emph{Expert Syst. Appl.}, vol. 233, p. 120946, 2023.

\bibitem{si2022target}
P.~Si, Z.~Fu, L.~Shu, Y.~Yang, K.~Huang, and Y.~Liu, ``Target-barrier coverage improvement in an insecticidal lamps internet of {UAV}s,'' \emph{IEEE Trans. Veh. Technol.}, vol.~71, no.~4, pp. 4373--4382, 2022.

\bibitem{chien2021UAV}
W.-C. Chien, M.~M. Hassan, A.~Alsanad, and G.~Fortino, ``{UAV}--assisted joint wireless power transfer and data collection mechanism for sustainable precision agriculture in 5{G},'' \emph{IEEE Micro}, vol.~42, no.~1, pp. 25--32, 2021.

\bibitem{zhang2021deploying}
C.~Zhang, M.~Dong, and K.~Ota, ``Deploying sdn control in {I}nternet of {UAV}s: {Q}-learning-based edge scheduling,'' \emph{IEEE Transactions on Network and Service Management}, vol.~18, no.~1, pp. 526--537, 2021.

\bibitem{chen2021identification}
C.-J. Chen, Y.-Y. Huang, Y.-S. Li, Y.-C. Chen, C.-Y. Chang, and Y.-M. Huang, ``Identification of fruit tree pests with deep learning on embedded drone to achieve accurate pesticide spraying,'' \emph{IEEE Access}, vol.~9, pp. 21\,986--21\,997, 2021.

\bibitem{sun2021aoi}
M.~Sun, X.~Xu, X.~Qin, and P.~Zhang, ``{A}o{I}-energy-aware {UAV}-assisted data collection for {I}o{T} networks: A deep reinforcement learning method,'' \emph{IEEE Internet Things J.}, vol.~8, no.~24, pp. 17\,275--17\,289, 2021.

\bibitem{yi2020deep}
M.~Yi, X.~Wang, J.~Liu, Y.~Zhang, and B.~Bai, ``Deep reinforcement learning for fresh data collection in {UAV}-assisted {I}o{T} networks,'' in \emph{IEEE Conf. Comput. Commun. Workshops (INFOCOM WKSHPS)}.\hskip 1em plus 0.5em minus 0.4em\relax IEEE, 2020, pp. 716--721.

\bibitem{feng2023age}
H.~Feng, J.~Wang, Z.~Fang, J.~Qian, and K.-C. Chen, ``Age of information in {UAV} aided wireless sensor networks relying on blockchain,'' \emph{IEEE Trans. Veh. Technol.}, 2023.

\bibitem{lu2021covertness}
X.~Lu, W.~Yang, S.~Yan, Z.~Li, and D.~W.~K. Ng, ``Covertness and timeliness of data collection in {UAV}-aided wireless-powered {I}o{T},'' \emph{IEEE Internet Things J.}, vol.~9, no.~14, pp. 12\,573--12\,587, 2021.

\bibitem{wang2019energy}
Z.~Wang, R.~Liu, Q.~Liu, J.~S. Thompson, and M.~Kadoch, ``Energy-efficient data collection and device positioning in {UAV}-assisted {I}o{T},'' \emph{IEEE Internet Things J.}, vol.~7, no.~2, pp. 1122--1139, 2019.

\bibitem{nazib2021energy}
R.~A. Nazib and S.~Moh, ``Energy-efficient and fast data collection in {UAV}-aided wireless sensor networks for hilly terrains,'' \emph{IEEE Access}, vol.~9, pp. 23\,168--23\,190, 2021.

\bibitem{hao2022joint}
C.~Hao, Y.~Chen, Z.~Mai, G.~Chen, and M.~Yang, ``Joint optimization on trajectory, transmission and time for effective data acquisition in {UAV}-enabled iot,'' \emph{IEEE Trans. Veh. Technol.}, vol.~71, no.~7, pp. 7371--7384, 2022.

\bibitem{zhan2019completion}
C.~Zhan and Y.~Zeng, ``Completion time minimization for multi-{UAV}-enabled data collection,'' \emph{IEEE Trans. Wirel. Commun.}, vol.~18, no.~10, pp. 4859--4872, 2019.

\bibitem{yan2020uav}
H.~Yan, Y.~Chen, and S.-H. Yang, ``{UAV}-enabled wireless power transfer with base station charging and {UAV} power consumption,'' \emph{IEEE Trans. Veh. Technol.}, vol.~69, no.~11, pp. 12\,883--12\,896, 2020.

\bibitem{fu2021joint}
Y.~Fu, H.~Mei, K.~Wang, and K.~Yang, ``Joint optimization of 3{D} trajectory and scheduling for solar-powered {UAV} systems,'' \emph{IEEE Trans. Veh. Technol.}, vol.~70, no.~4, pp. 3972--3977, 2021.

\bibitem{li2019rechargeable}
X.~Li, H.~Yao, J.~Wang, S.~Wu, C.~Jiang, and Y.~Qian, ``Rechargeable multi-{UAV} aided seamless coverage for {Q}o{S}-guaranteed {I}o{T} networks,'' \emph{IEEE Internet Things J.}, vol.~6, no.~6, pp. 10\,902--10\,914, 2019.

\bibitem{luo2020joint}
W.~Luo, Y.~Shen, B.~Yang, S.~Wang, and X.~Guan, ``Joint 3-{D} trajectory and resource optimization in multi-{UAV}-enabled {I}o{T} networks with wireless power transfer,'' \emph{IEEE Internet Things J.}, vol.~8, no.~10, pp. 7833--7848, 2020.

\bibitem{fan2023channel}
X.~Fan, H.~Zhou, K.~Sun, X.~Chen, and N.~Wang, ``Channel assignment and power allocation utilizing {NOMA} in long-distance {UAV} wireless communication,'' \emph{IEEE Trans. Veh. Technol.}, 2023.

\bibitem{wang2023secrecy}
S.~Wang, L.~Li, R.~Ruby, X.~Ruan, J.~Zhang, and Y.~Zhang, ``Secrecy-energy-efficiency {UAV}-enabled two-way maximization for relay systems,'' \emph{IEEE Trans. Veh. Technol.}, 2023.

\bibitem{wang2023joint}
J.~Wang, X.~Zhou, H.~Zhang, and D.~Yuan, ``Joint trajectory design and power allocation for {UAV} assisted network with user mobility,'' \emph{IEEE Trans. Veh. Technol.}, 2023.

\bibitem{yu2023joint}
Y.~Yu, X.~Liu, Z.~Liu, and T.~S. Durrani, ``Joint trajectory and resource optimization for ris assisted {UAV} cognitive radio,'' \emph{IEEE Trans. Veh. Technol.}, 2023.

\bibitem{wu2023uav}
L.~Wu, W.~Wang, Z.~Ji, Y.~Yang, K.~Cumanan, G.~Chen, Z.~Ding, and O.~A. Dobre, ``Uav-assisted maritime legitimate surveillance: Joint trajectory design and power allocation,'' \emph{IEEE Trans. Veh. Technol.}, 2023.

\bibitem{nguyen2021uav}
P.~X. Nguyen, V.-D. Nguyen, H.~V. Nguyen, and O.-S. Shin, ``Uav-assisted secure communications in terrestrial cognitive radio networks: Joint power control and 3d trajectory optimization,'' \emph{IEEE Trans. Veh. Technol.}, vol.~70, no.~4, pp. 3298--3313, 2021.

\bibitem{liu2022energy}
Z.~Liu, X.~Liu, V.~C. Leung, and T.~S. Durrani, ``Energy-efficient resource allocation for dual-{NOMA}-{UAV} assisted {I}nternet of {T}hings,'' \emph{IEEE Trans. Veh. Technol.}, vol.~72, no.~3, pp. 3532--3543, 2022.

\bibitem{al2014optimal}
A.~Al-Hourani, S.~Kandeepan, and S.~Lardner, ``Optimal {LAP} altitude for maximum coverage,'' \emph{IEEE Wirel. Commun. Lett.}, vol.~3, no.~6, pp. 569--572, 2014.

\bibitem{you2020hybrid}
C.~You and R.~Zhang, ``Hybrid offline-online design for {UAV}-enabled data harvesting in probabilistic {L}o{S} channels,'' \emph{IEEE Trans. Wirel. Commun.}, vol.~19, no.~6, pp. 3753--3768, 2020.

\bibitem{zhan2020completion}
C.~Zhan, H.~Hu, X.~Sui, Z.~Liu, and D.~Niyato, ``Completion time and energy optimization in the {UAV}-enabled mobile-edge computing system,'' \emph{IEEE Internet Things J.}, vol.~7, no.~8, pp. 7808--7822, 2020.

\bibitem{zeng2019energy}
Y.~Zeng, J.~Xu, and R.~Zhang, ``Energy minimization for wireless communication with rotary-wing {UAV},'' \emph{IEEE Trans. Wirel. Commun.}, vol.~18, no.~4, pp. 2329--2345, 2019.

\bibitem{wu2018joint}
Q.~Wu, Y.~Zeng, and R.~Zhang, ``Joint trajectory and communication design for multi-{UAV} enabled wireless networks,'' \emph{IEEE Trans. Wirel. Commun.}, vol.~17, no.~3, pp. 2109--2121, 2018.

\bibitem{samir2019uav}
M.~Samir, S.~Sharafeddine, C.~M. Assi, T.~M. Nguyen, and A.~Ghrayeb, ``{UAV} trajectory planning for data collection from time-constrained {I}o{T} devices,'' \emph{IEEE Trans. Wirel. Commun.}, vol.~19, no.~1, pp. 34--46, 2019.

\bibitem{vansteenwegen2011orienteering}
P.~Vansteenwegen, W.~Souffriau, and D.~Van~Oudheusden, ``The orienteering problem: A survey,'' \emph{Eur. J. Oper. Res.}, vol. 209, no.~1, pp. 1--10, 2011.

\bibitem{zhan2018trajectory}
C.~Zhan, Y.~Zeng, and R.~Zhang, ``Trajectory design for distributed estimation in {UAV}-enabled wireless sensor network,'' \emph{IEEE Trans. Veh. Technol.}, vol.~67, no.~10, pp. 10\,155--10\,159, 2018.

\bibitem{kumawat2017multi}
I.~R. Kumawat, S.~J. Nanda, and R.~K. Maddila, ``Multi-objective whale optimization,'' in \emph{Tencon 2017-2017 ieee region 10 conference}.\hskip 1em plus 0.5em minus 0.4em\relax IEEE, 2017, pp. 2747--2752.

\bibitem{azizi2022multiobjective}
M.~Azizi, S.~Talatahari, N.~Khodadadi, and P.~Sareh, ``Multiobjective atomic orbital search ({MOAOS}) for global and engineering design optimization,'' \emph{IEEE Access}, vol.~10, pp. 67\,727--67\,746, 2022.

\bibitem{khodadadi2021multi}
N.~Khodadadi, M.~Azizi, S.~Talatahari, and P.~Sareh, ``Multi-objective crystal structure algorithm ({MOC}ry{S}t{A}l): Introduction and performance evaluation,'' \emph{IEEE Access}, vol.~9, pp. 117\,795--117\,812, 2021.

\bibitem{ramadan2022accurate}
A.~Ramadan, S.~Kamel, M.~H. Hassan, E.~M. Ahmed, and H.~M. Hasanien, ``Accurate photovoltaic models based on an adaptive opposition artificial hummingbird algorithm,'' \emph{Electronics}, vol.~11, no.~3, p. 318, 2022.

\bibitem{bhagat2023application}
S.~K. Bhagat, L.~C. Saikia, and N.~R. Babu, ``Application of artificial hummingbird algorithm in a renewable energy source integrated multi-area power system considering {F}uzzy based tilt integral derivative controller,'' \emph{e-Prime-Advances in Electrical Engineering, Electronics and Energy}, vol.~4, p. 100153, 2023.

\bibitem{chen2023pso}
K.~Chen, L.~Chen, and G.~Hu, ``{PSO}-incorporated hybrid artificial hummingbird algorithm with elite opposition-based learning and {C}auchy mutation: A case study of shape optimization for {CSGC}--ball curves,'' \emph{Biomimetics}, vol.~8, no.~4, p. 377, 2023.

\bibitem{zhao2022effective}
W.~Zhao, Z.~Zhang, S.~Mirjalili, L.~Wang, N.~Khodadadi, and S.~M. Mirjalili, ``An effective multi-objective artificial hummingbird algorithm with dynamic elimination-based crowding distance for solving engineering design problems,'' \emph{Comput Methods Appl Mech Eng.}, vol. 398, p. 115223, 2022.

\bibitem{deb2002fast}
K.~Deb, A.~Pratap, S.~Agarwal, and T.~Meyarivan, ``A fast and elitist multiobjective genetic algorithm: {NSGA-II},'' \emph{IEEE transactions on evolutionary computation}, vol.~6, no.~2, pp. 182--197, 2002.

\bibitem{zhao2022artificial}
W.~Zhao, L.~Wang, and S.~Mirjalili, ``Artificial hummingbird algorithm: A new bio-inspired optimizer with its engineering applications,'' \emph{Comput Methods Appl Mech Eng.}, vol. 388, p. 114194, 2022.

\bibitem{yuan2008hydrothermal}
X.~Yuan, B.~Cao, B.~Yang, and Y.~Yuan, ``Hydrothermal scheduling using chaotic hybrid differential evolution,'' \emph{Energy Convers. Manag.}, vol.~49, no.~12, pp. 3627--3633, 2008.

\bibitem{saremi2014biogeography}
S.~Saremi, S.~Mirjalili, and A.~Lewis, ``Biogeography-based optimisation with chaos,'' \emph{Neural Comput. Appl.}, vol.~25, no.~5, pp. 1077--1097, 2014.

\bibitem{liu2022multiobjective}
L.~Liu, A.~Wang, G.~Sun, and J.~Li, ``Multiobjective optimization for improving throughput and energy efficiency in {UAV}-enabled {I}o{T},'' \emph{IEEE Internet Things J.}, vol.~9, no.~20, pp. 20\,763--20\,777, 2022.

\bibitem{ali2011improving}
M.~Ali and M.~Pant, ``Improving the performance of differential evolution algorithm using {C}auchy mutation,'' \emph{Soft Comput.}, vol.~15, no.~5, pp. 991--1007, 2011.

\bibitem{wang2016opposition}
G.-G. Wang, S.~Deb, A.~H. Gandomi, and A.~H. Alavi, ``Opposition-based krill herd algorithm with {C}auchy mutation and position clamping,'' \emph{Neurocomputing}, vol. 177, pp. 147--157, 2016.

\bibitem{zhao2021enhanced}
S.~Zhao, P.~Wang, A.~A. Heidari, X.~Zhao, C.~Ma, and H.~Chen, ``An enhanced {C}auchy mutation grasshopper optimization with trigonometric substitution: Engineering design and feature selection,'' \emph{Eng. Comput.}, pp. 1--34, 2021.

\bibitem{khodadadi2022moavoa}
N.~Khodadadi, F.~Soleimanian~Gharehchopogh, and S.~Mirjalili, ``{MOAVOA}: A new multi-objective artificial vultures optimization algorithm,'' \emph{Neural Comput. Appl.}, vol.~34, no.~23, pp. 20\,791--20\,829, 2022.

\end{thebibliography}
\clearpage
\begin{IEEEbiography}[{\includegraphics[width=1in,height=1.25in,clip,keepaspectratio]{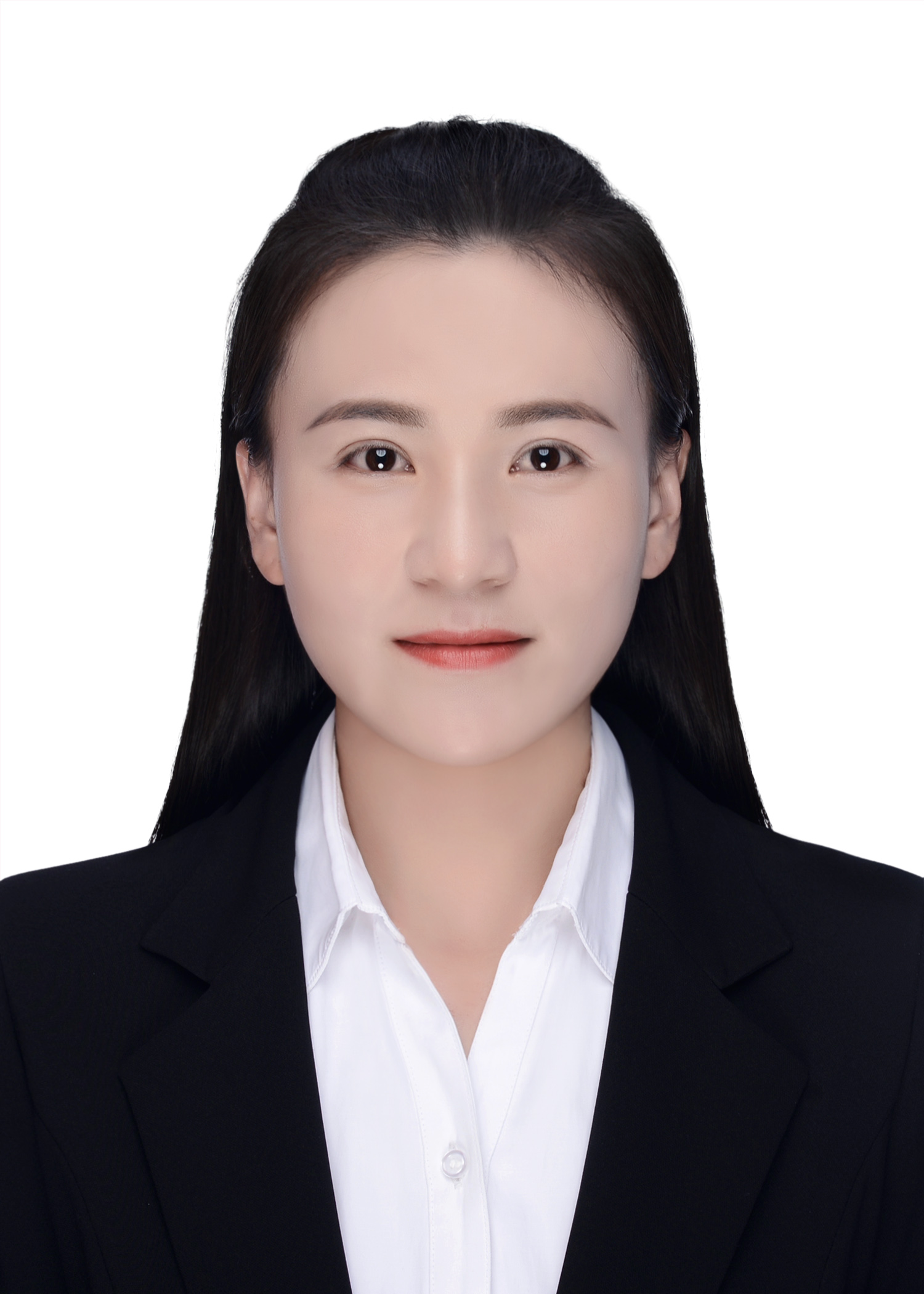}}]{Lingling~Liu} received the BS degree in Software Engineering from Jinzhong University in 2018. Currently, she is currently pursuing the Ph.D. degree in computer science at Jilin University. Her research interests include wireless sensor networks and antenna array optimization.
\end{IEEEbiography}

\begin{IEEEbiography}[{\includegraphics[width=1in,height=1.25in,clip,keepaspectratio]{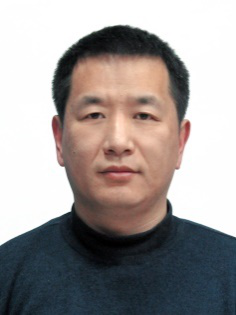}}]{Aimin~Wang} received the B.S. degree in Computer Software and M.S. degree in Computer Application Technology from Jilin University. He also received the Ph.D. degree in Communication and Information Systems from Jilin University. He is currently an associate professor at Jilin University. His research interests are wireless sensor networks and QoS for multimedia transmission.
\end{IEEEbiography}

\begin{IEEEbiography}[{\includegraphics[width=1in,height=1.25in,clip,keepaspectratio]{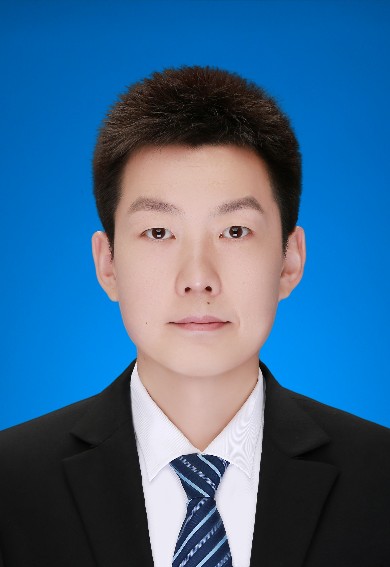}}]{Geng~Sun} received a B.S. degree in Communication Engineering from Dalian Polytechnic University, China, and the Ph.D. degree in Computer Science from Jilin University, China, in 2011 and 2018, respectively. He was a visiting researcher in the School of Electrical and Computer Engineering at Georgia Institute of Technology, USA. He is currently an Associate Professor in College of Computer Science and Technology at Jilin University, and His research interests include wireless sensor networks, antenna array, collaborative beamforming and optimizations.
\end{IEEEbiography}

\begin{IEEEbiography}[{\includegraphics[width=1in,height=1.25in,clip,keepaspectratio]{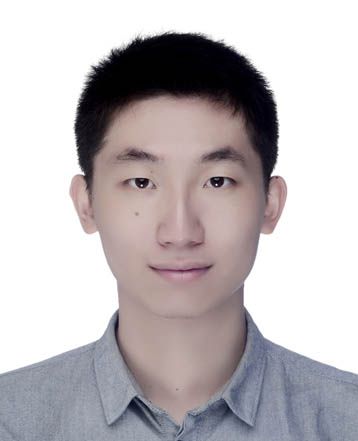}}]{Jiahui~Li} received a B.S. degree in software engineering from Jilin University, Changchun, China, in 2018. Currently, he is studying for the M.Sc. degree in Computer Science and Technology from Jilin University, Changchun, China. His current research focuses on UAV networks, antenna arrays, and optimization.
\end{IEEEbiography}

\begin{IEEEbiography}[{\includegraphics[width=1in,height=1.25in,clip,keepaspectratio]{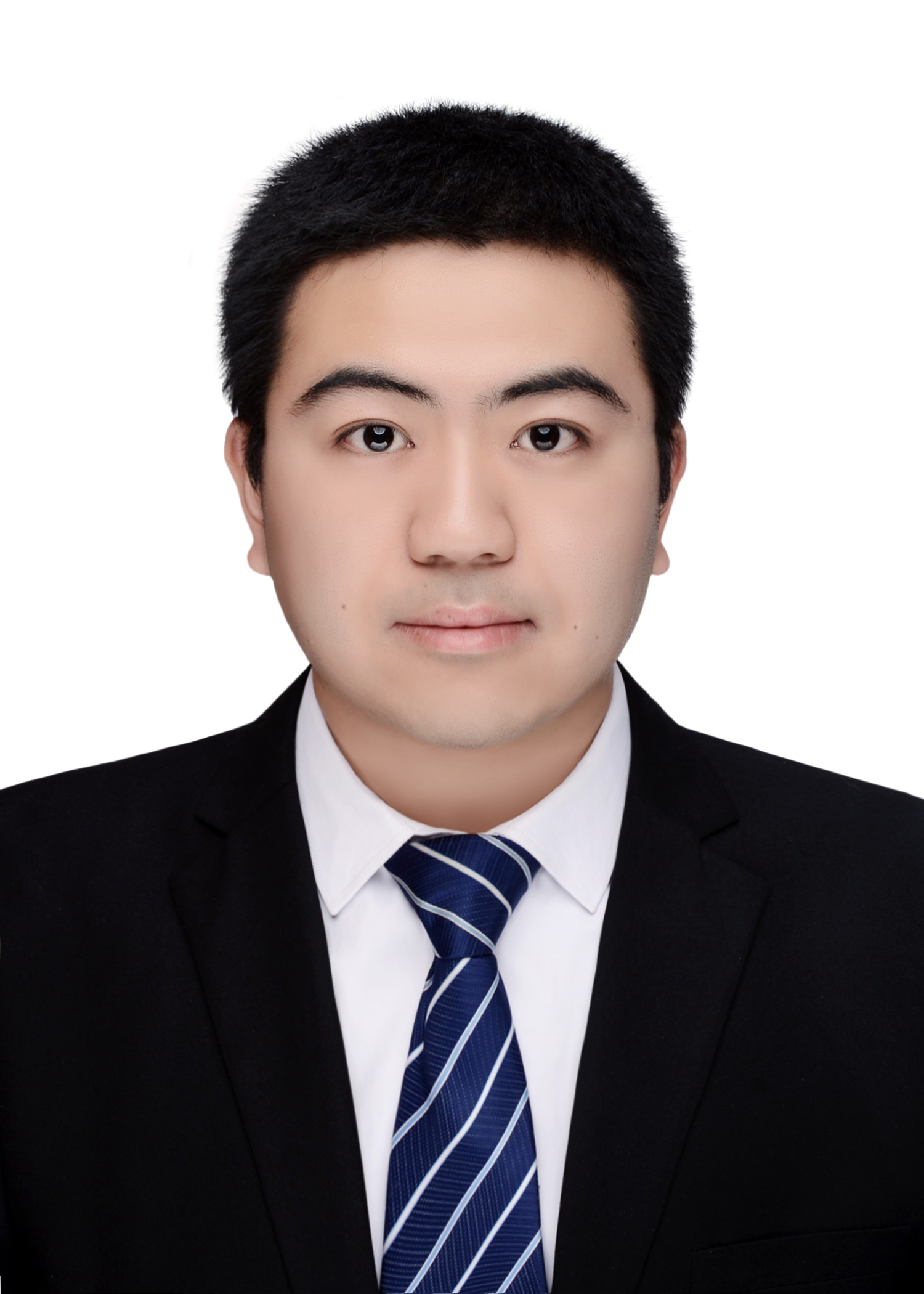}}]{Hongyang Pan} received the B.S. degree in process equipment and control engineering from Dalian University of Technology in 2017. He is currently pursuing the Ph.D. degree in computer science and technology, Jilin University. His research interests include the wireless communications and optimizations.
\end{IEEEbiography}

\begin{IEEEbiography}[{\includegraphics[width=1in,height=1.25in,keepaspectratio]{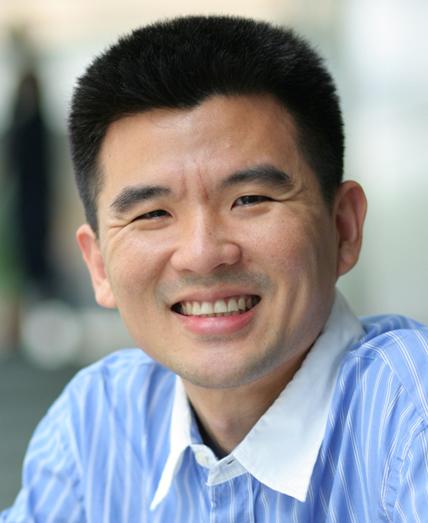}}]
{Tony Q.S. Quek}(S'98-M'08-SM'12-F'18) received the B.E.\ and M.E.\ degrees in electrical and electronics engineering from the Tokyo Institute of Technology in 1998 and 2000, respectively, and the Ph.D.\ degree in electrical engineering and computer science from the Massachusetts Institute of Technology in 2008. Currently, he is the Cheng Tsang Man Chair Professor with Singapore University of Technology and Design (SUTD) and ST Engineering Distinguished Professor. He also serves as the Director of the Future Communications R\&D Programme, the Head of ISTD Pillar, and the Deputy Director of the SUTD-ZJU IDEA. His current research topics include wireless communications and networking, network intelligence, non-terrestrial networks, open radio access network, and 6G.
	
Dr.\ Quek has been actively involved in organizing and chairing sessions, and has served as a member of the Technical Program Committee as well as symposium chairs in a number of international conferences. He is currently serving as an Area Editor for the {\scshape IEEE Transactions on Wireless Communications}. 
	
Dr.\ Quek was honored with the 2008 Philip Yeo Prize for Outstanding Achievement in Research, the 2012 IEEE William R. Bennett Prize, the 2015 SUTD Outstanding Education Awards -- Excellence in Research, the 2016 IEEE Signal Processing Society Young Author Best Paper Award, the 2017 CTTC Early Achievement Award, the 2017 IEEE ComSoc AP Outstanding Paper Award, the 2020 IEEE Communications Society Young Author Best Paper Award, the 2020 IEEE Stephen O. Rice Prize, the 2020 Nokia Visiting Professor, and the 2022 IEEE Signal Processing Society Best Paper Award. He is a Fellow of IEEE and a Fellow of the Academy of Engineering Singapore.
\end{IEEEbiography}

\end{document}